\PassOptionsToPackage{pdfpagelabels=false}{hyperref} 
\documentclass[usenatbib,useAMS]{mnras}

\newcommand\lsim{\mathrel{\rlap{\lower4pt\hbox{\hskip1pt$\sim$}}
        \raise1pt\hbox{$<$}}}
\newcommand\gsim{\mathrel{\rlap{\lower4pt\hbox{\hskip1pt$\sim$}}
        \raise1pt\hbox{$>$}}}

\usepackage{times}
\usepackage{graphicx}
\usepackage{verbatim}
\usepackage{tikz}
\usepackage{color}

\usepackage[T1]{fontenc}
\usepackage{aecompl}
\usepackage{calligra}
\DeclareMathAlphabet{\mathcalligra}{T1}{calligra}{m}{n}
\DeclareFontShape{T1}{calligra}{m}{n}{<->s*[2.2]callig15}{}

\usepackage{hyperref}
\hypersetup{
    pdfnewwindow=true,      
    colorlinks=true,       
    linkcolor=blue,          
    citecolor=cyan,        
    filecolor=blue,      
    urlcolor=blue           
}

\def\prd{PRD}
\def\apj{ApJ}                 
\def\apjl{ApJL}               
\def\mnras{MNRAS}             
\def\aap{A\&A}                
\def\aaps{A\&AS}              
\def\nat{Nature}              
\def\araa{ARA\&A}             
\def\ssr{SSR}                 

\def\eg{\textit{e.g.}}

\def\bin{\rm{bin}}

\def\out{\rm{out}}

\def\min{\rm{min}}
\def\max{\rm{max}}

\def\Msun{{M_{\odot}}}

\begin{document}

\title[MBHB self-lensing]{Periodic self-lensing from accreting massive black hole binaries}
\author[D. J. D'Orazio, R. Di Stefano]{Daniel J. D'Orazio$^1$\thanks{daniel.dorazio@cfa.harvard.edu; rdistefano@cfa.harvard.edu},
  Rosanne Di Stefano$^1$ \\
     $^1$Astronomy Department, Harvard University, 60 Garden Street, Cambridge, MA 02138}

\maketitle
\begin{abstract}
Nearly $150$ massive black hole binary (MBHB) candidates at sub-pc orbital
separations have been reported in recent literature. Nevertheless, the
definitive detection of even a single such object remains elusive.  If at
least one of the black holes is accreting, the light emitted from its
accretion disc will be lensed by the other black hole for binary orbital
inclinations near to the line of sight. This binary self-lensing could provide
a unique signature of compact MBHB systems. We show that, for MBHBs with
masses in the range $10^6 - 10^{10} \Msun$ and with orbital periods less than
$\sim 10$ yr, strong lensing events should occur in one to 10s of percent of
MBHB systems that are monitored for an entire orbit. Lensing events will last
from days for the less massive, shorter period MBHBs up to a year for the most
massive $\sim 10$ year orbital period MBHBs. At small inclinations of the
binary orbit to the line of sight, lensing must occur and will be accompanied
by periodicity due to the relativistic Doppler boost. Flares at the same phase
as the otherwise average flux of the Doppler modulation would be a smoking gun
signature of self-lensing and can be used to constrain binary parameters. For
MBHBs with separation $\gsim 100$ Schwarzschild radii, we show that 
finite-sized source effects could serve as a probe of MBH accretion disc structure.
Finally, we stress that our lensing probability estimate implies that $\sim10$
of the known MBHB candidates identified through quasar periodicity should
exhibit strong lensing flares.
\end{abstract}


\maketitle

\section{Introduction}

The merger of two galaxies, each containing a massive black hole (MBH), can
yield a compact, massive black hole binary \citep[MBHB;][]{kr95, KH2000, ff05,
KormendyHo2013}. The MBHs may merge, producing gravitational radiation
\citep[][]{Begel:Blan:Rees:1980}. Discoveries of MBHBs therefore allow us to
learn about past galaxy interactions and to predict important future events.
Such systems can be identified electromagnetically only if matter falling on to
the MBH(s) emits light. In recent years, active galactic nuclei (AGN) have
been examined for evidence that the accreting MBH that powers it may be in a
binary. Most relevant to this study, recent time-domain surveys have selected
MBHB candidates by identifying periodically varying optical emission from such
an AGN \citep{Graham+2015a, Graham+2015b, Charisi+2016}. The periodicity could
be caused by variable accretion on to a binary \citep{Hayasaki:2007,
MacFadyen:2008, Cuadra:2009, Roedig:2012:Trqs, Noble+2012, ShiKrolik:2012,
DHM:2013:MNRAS, Farris:2014, Gold:GRMHD_CBD:2014, Gold:GRMHD_CBDII:2014,
Dunhill+2015, ShiKrolik:2015, Farris:2015:Cool, PG1302MNRAS:2015a,
D'Orazio:CBDTrans:2016, MunozLai:2016}, or relativistic Doppler boost of
emission emanating from gas bound to the orbiting BHs
\citep{PG1302Nature:2015b}.

In this paper we explore a signature which can help to discover more MBHB
candidates, vet the candidacy of systems identified in other ways, measure the
binary's orbital parameters, and help to map the accretion geometry.
Specifically, we consider the gravitational lensing of the accretion flow of
one BH by its companion. Binary self-lensing is a phenomenon that must occur
in systems with favourable orientations \citep[see also][]{Rahvar+2011}. We
will show that self-lensing is expected in a non-negligible number (a few to
10s of percent depending on binary parameters) of accreting MBHBs.

We compute the lensing effects and show that they can be detected and
correctly interpreted. The signatures of self-lensing are most dramatic if the
MBHB has mass near or above $10^6 \Msun$ and a small orbital inclination to
the line of sight. Lensing produces a flare at specific values of the orbital
phase. The duration of this flare is proportional to the square root of the
mass of the lensing BH and can range from days to years for MBHBs depending on
the binary orbital period, mass, and mass ratio.

Small binary orbital inclinations also favour detection of periodicity
generated by the relativistic Doppler boost. If the accretion disc emission is
observed to be modulated in time due to the relativistic Doppler effect, then
the lensing signatures must occur at a specific phase in the Doppler-boost
light curve, namely at the times of average brightness when one BH is behind
the other, moving transverse to the line of sight. Hence, for example, an
observation of lensing flares at what would otherwise be the flux averages of
a nearly sinusoidal light curve would provide strong evidence for a MBHB and
for the Doppler-boost interpretation.

We show that, depending on the binary separation and masses of the BHs, the
accretion disc source can act either as a point source or a finite-sized
source. In the point-source case, a unique identifier of the lensing event is
its appearance in all wavelengths, because gravitational lensing is
achromatic. We discuss in \S\ref{S:Discussion}, however, the practical
difficulties that may be associated with determining achromaticity in a real
MBHB system. Assuming that the disc size is proportional to the Hill radius,
finite-sized source lensing will become important for larger separation
binaries and cause wavelength-dependent lensing flares. Wavelength dependence
arises because the accretion disc will emit at different wavelengths at
different radii; hence, a lensing event of an accretion disc in the  
finite-sized source case will probe the accretion flow around the lensed BHs. In
cases where there are two discs, one for each BH, the achromaticity of lensing
will allow us to distinguish the spectra of the individual discs and possibly
the circumbinary disc.

Finally, once the existence of a specific MBHB for which we observe
gravitational lensing is firmly established, any future changes in the
emission from the BH can be studied in more detail than would otherwise be
possible through the magnification effects of the `telescopes' provided by
the BHs themselves.

\section{Lensing Calculations}

\subsection{System scales}
\label{S:System scales}

We envision the galactic merger process delivering MBHs and gas to the centre
of a newly formed galaxy \citep{BH1992, Barnes:1996, Barnes:2002,
Mayer:2013:MBHBGasRev}. When the MBHBs harden sufficiently (sub-pc
separations), they can be surrounded by a circumbinary gaseous disc. This disc
feeds `mini-discs', which are bound to, and feed each individual MBH. Such gas
accretion can generate bright emission, as in the case of a single MBH, but
modulated at the binary orbital period and its multiples
\citep[\eg,][]{DHM:2013:MNRAS, Farris:2014}. Gas may also be instrumental in
deciphering the so-called final parsec problem \citep{Begel:Blan:Rees:1980,
Milosavljevic:2003:FPcP, ArmNat:2005}, bringing MBHBs down to the orbital
separations considered here. We consider such accreting binary systems and the
possibility that emission from the gas bound to one MBH can be gravitationally
lensed by the other, generating a detectable flare that may act as a unique
identifier of the MBHB and disc system.

The range of binary orbital parameters of interest to this study are set
primarily by observational constraints. We set a maximum binary separation by
requiring that the entire orbital period of the binary be observed twice
within the lifetime of a survey. While, as we discuss in \S
\ref{S:Discussion}, the detection of a single lensing flare may be sufficient
for determining its origin, to be conservative, we require two successive
observations of the lensing flare, allowing a minimal identification of
periodicity. This motivates us to set an upper limit of $\sim 10$ years on the
observed binary orbital period, corresponding to binary separations of $a
\lsim 0.01  \rm{pc} \ M^{1/3}_8 (1+z)^{-2/3}$, where $z$ is the redshift of
the MBHB host galaxy, $M_8$ is the total binary mass in units of $10^8 \Msun$,
and we assume that the binary is on a circular orbit throughout
(however, see \S~\ref{S:Extensions}). This upper limit on the binary
separation is always within the approximate size of a gravitationally stable,
circumbinary disc \citep{Goodman:2003, HKM09}.

While we set no minimum binary period, we note that, observationally, a minimum
will be set by the sample size of quasars that can be monitored in a survey.
MBHBs with shorter orbital periods have shorter lifetimes and hence have a
lower probability of detection \cite[\eg,][]{HKM09}\footnote{ 
Combining the models of \citep{HKM09} for gas plus gravitational wave driven
orbital decay and an AGN sample size of $\sim10^5$ \citep[\eg, the Catalina
Real Time Transient Survey;][]{CRTS1:Drake:2009,
CRTS3:2011Mahabal, CRTS4:Djorgovski:2011}, we estimate the binary period below
which less than one MBHB is expected to be found. This minimum detectable MBHB
period is $P_{\min} \approx 79 \ \rm{days} \ M^{5/8}_8 q^{3/8}_s$, or $a \geq
8 \times 10^{-4} \rm{pc} \ M^{3/4}_8 q^{1/4}_s$.}.
On the theoretical side, we treat the lens as a point mass and carry out all
lensing calculations in the weak-field gravity limit. For very close binaries,
however, one must take into account strong-field gravitational effects.

\subsection{Lensing scales} 

\begin{figure}
\begin{center}
\includegraphics[scale=0.33]{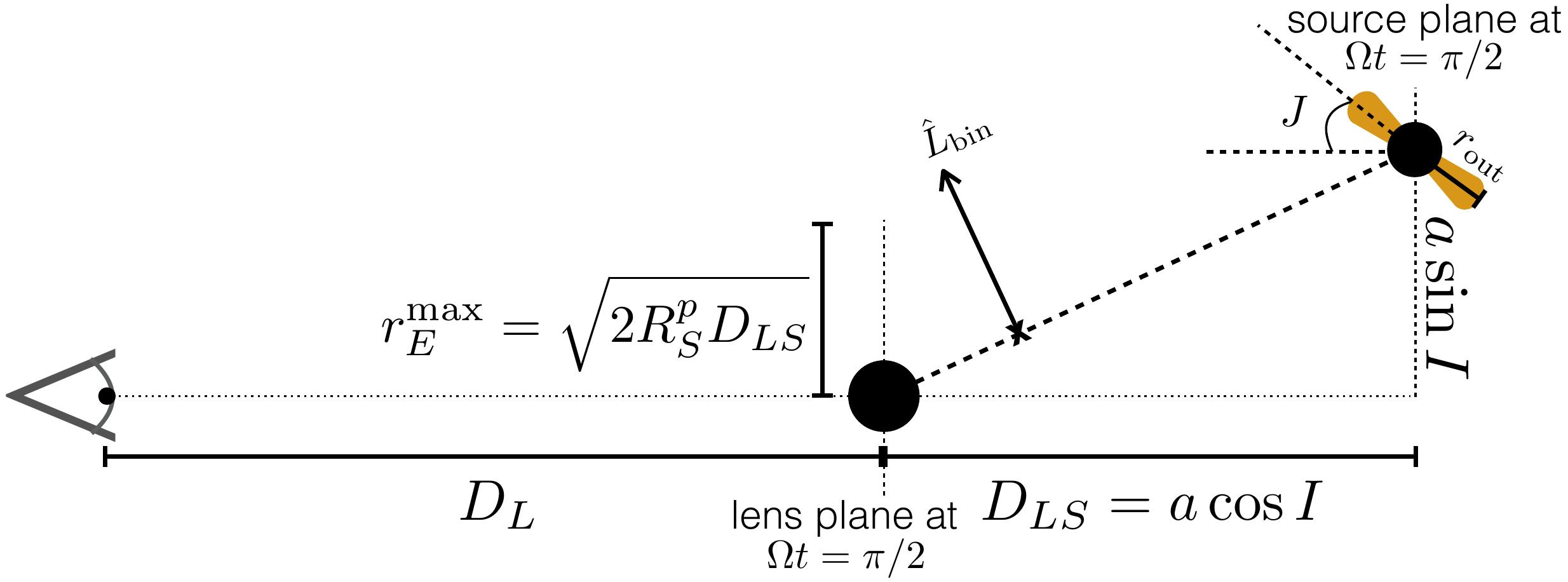}
\end{center}
\caption{ 
The lensing geometry at maximum magnification for a given binary inclination
$I$ and disc inclination $J$: viewed as the light emitting secondary BH passes
behind the primary BH at a closest angular separation of $a\sin{I}/D_{L}$, at
time $\Omega t = \pi/2$. Also labeled is the outer radius of the secondary
accretion disc $r_{\rm{out}}$ and the Einstein radius of the primary BH at
this moment in the orbit, $r^{\max}_E$, where $R^p_S$ is the Schwarzschild
radius of the primary. $\hat{L}_{\bin}$ is the binary angular momentum unit
vector.
}
\label{Fig:Schem}
\end{figure}

Consider a binary with orbital period $P$, primary and secondary masses $M_p$
and $M_s$; $M_s \leq M_p$, total mass $M\equiv M_p+M_s$, and mass ratio $q
\equiv M_s/M_p$, where the subscripts $p$ and $s$ refer to the primary and
secondary BH respectively throughout. We study lensing of an emission region
bound to the secondary BH by the primary BH, or vice versa. The lensing
geometry for a snapshot of the binary orbit is depicted in Figure
\ref{Fig:Schem}.

Throughout we treat the lens as a point mass (we address this approximation
in \S\ref{S:Strong-field effects}). Whether the accretion disc may be treated
as a point source depends on its physical size relative to the Einstein radius
of the lens. The Einstein radius of the primary, when lensing the secondary,
is,
\begin{equation}
r^p_E = \sqrt{2 R^p_S a \cos{I} \sin{\Omega t}},
\label{Eq:rsE}
\end{equation}
where $R^p_S$ is the Schwarzschild radius of the primary, $I$ is the 
inclination of the binary plane to the line of sight,
$\Omega=2 \pi/P$ is the binary orbital angular frequency, $a$ is the binary
orbital separation given by $\Omega$ and $M$, and the Einstein radius $r^p_E$
is complex when the secondary passes in front of the primary. We have assumed
that the distances to the source and to the lens $D_L$ are equivalent since
$D_L \gg a$. 

From tidal truncation theory we know that the accretion flow around the
secondary BH can be as large in extent as $\eta \left( q/3 \right)^{1/3}a$,
where $\eta\leq 1$. Because the Einstein radius grows more slowly with
separation, as $\sqrt{a}$, the projected size of the secondary accretion flow
can be larger than the Einstein radius of the primary and, hence, for larger
separation (longer period) binaries, we must treat the source as finite in
size. We treat both the point source and finite-sized source cases below.

\subsection{Point Source}
\label{S:Point Source}

We first compute the lensing probability, time-scale, and time-dependent
magnification of the source accretion disc, treating it as a point source.
This is accurate for the highest energy emission that emanates from the
accretion disc inner edge. The point-source case also serves as a comparison
to the more general, finite-sized source case.

A significant lensing event occurs when the source passes within one
Einstein radius of the lens. Hence, the probability of observing a lensing
event, after observing the binary for an entire orbit, is the ratio of binary
inclination angles for which the source falls within one Einstein radius of the
lens to the total possible range of inclinations,
\begin{eqnarray}
\mathcal{P}_{s} &\sim& \frac{2}{\pi} \sin^{-1}\left(  \frac{r^{\max}_E}{a}   \right) \approx 2 \frac{\mathcal{T}_{s}}{P}, \nonumber \\
\mathcal{P}_{p} &=& q^{1/2} \mathcal{P}_{s},
\label{Eq:Prob}
\end{eqnarray}
where the subscripts $s$ and $p$ refer to the binary component that is being
lensed, $r^{\max}_E$ is the Einstein radius when the source is directly behind
the primary ($I=0$ and $\Omega t = \pi/2$ for the primary Einstein radius) and
$n_a \equiv a/R_S$ is the binary separation in units of  the Schwarzschild
radius of the total binary mass. In deriving Eq. (\ref{Eq:Prob})
we have neglected a factor of $\left(\cos{I}\right)^{-1/2}$, which is 
near-unity for the cases considered here. Note that when $n_a\rightarrow 2$,
emission from the secondary is strongly lensed at any binary inclination
(though we would no longer be in the weak-field regime and relativistic ray
tracing must be carried out).

We estimate the time-scale of a lensing event by calculating the fraction of
an orbit for which the source lies within the Einstein radius of the lens.
Assuming that the binary is inclined close enough to the line of sight for
significant lensing to occur,
\begin{eqnarray}
\mathcal{T}_{s} &\sim&  \frac{1}{\pi} \sin^{-1}{\left(\frac{r^{\max}_E}{a}\right)}  P  
\approx  \frac{1}{\pi} \sqrt{\frac{2}{n_a} (1+q)^{-1}} P,  \\ \nonumber
\mathcal{T}_{p} &=& q^{1/2} \mathcal{T}_{s},
\label{Eq:TSL}
\end{eqnarray}
where again the subscripts $s$ and $p$ refer to the binary component that is
being lensed. It is interesting to note that in the $n_a \rightarrow 2$ limit
mentioned above, the source is strongly lensed for the entire half-orbital
period for which the source is behind the lens.

\begin{figure*}
\begin{center}$
\begin{array}{c c}
\includegraphics[scale=0.3]{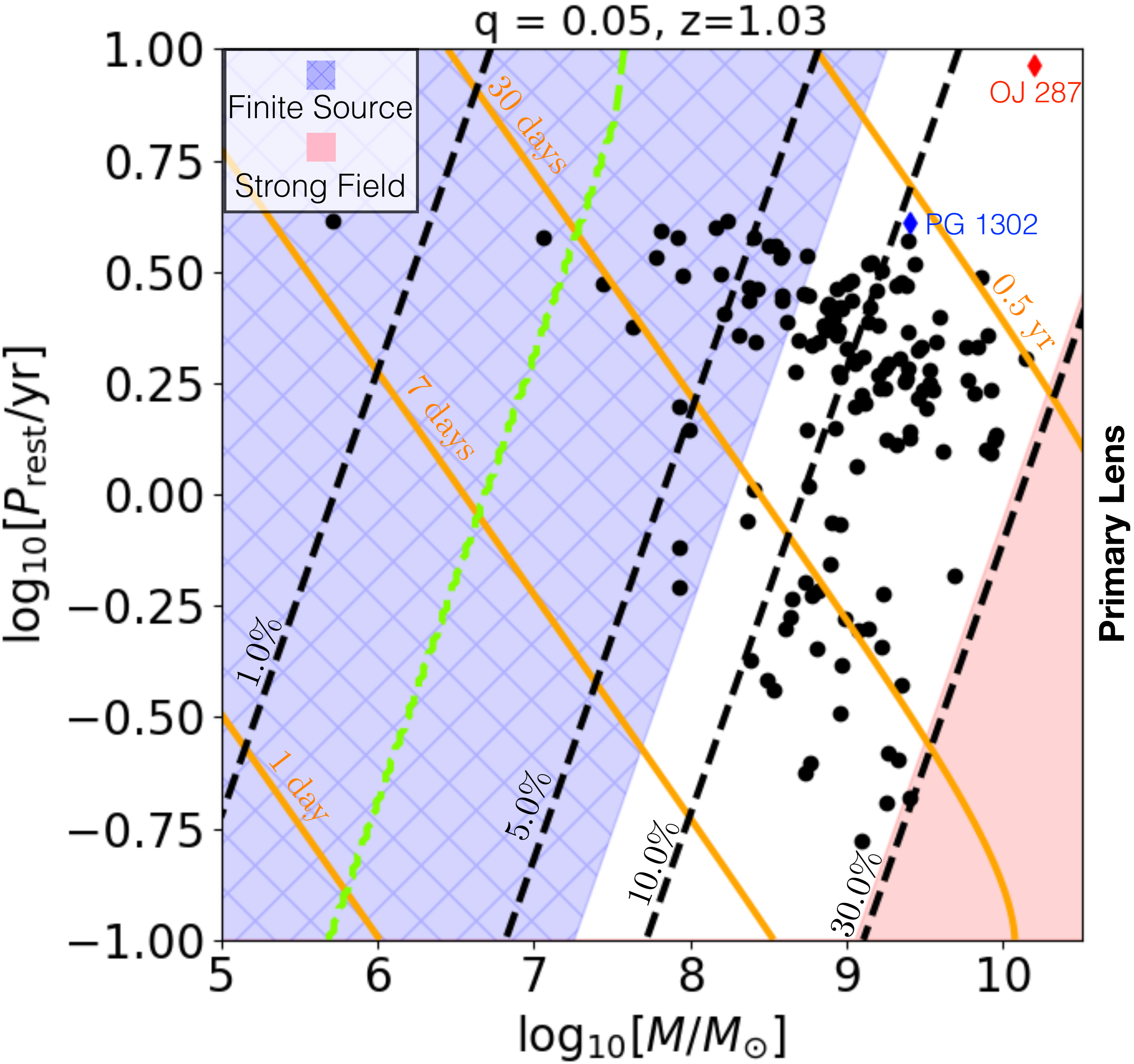} &
\includegraphics[scale=0.3]{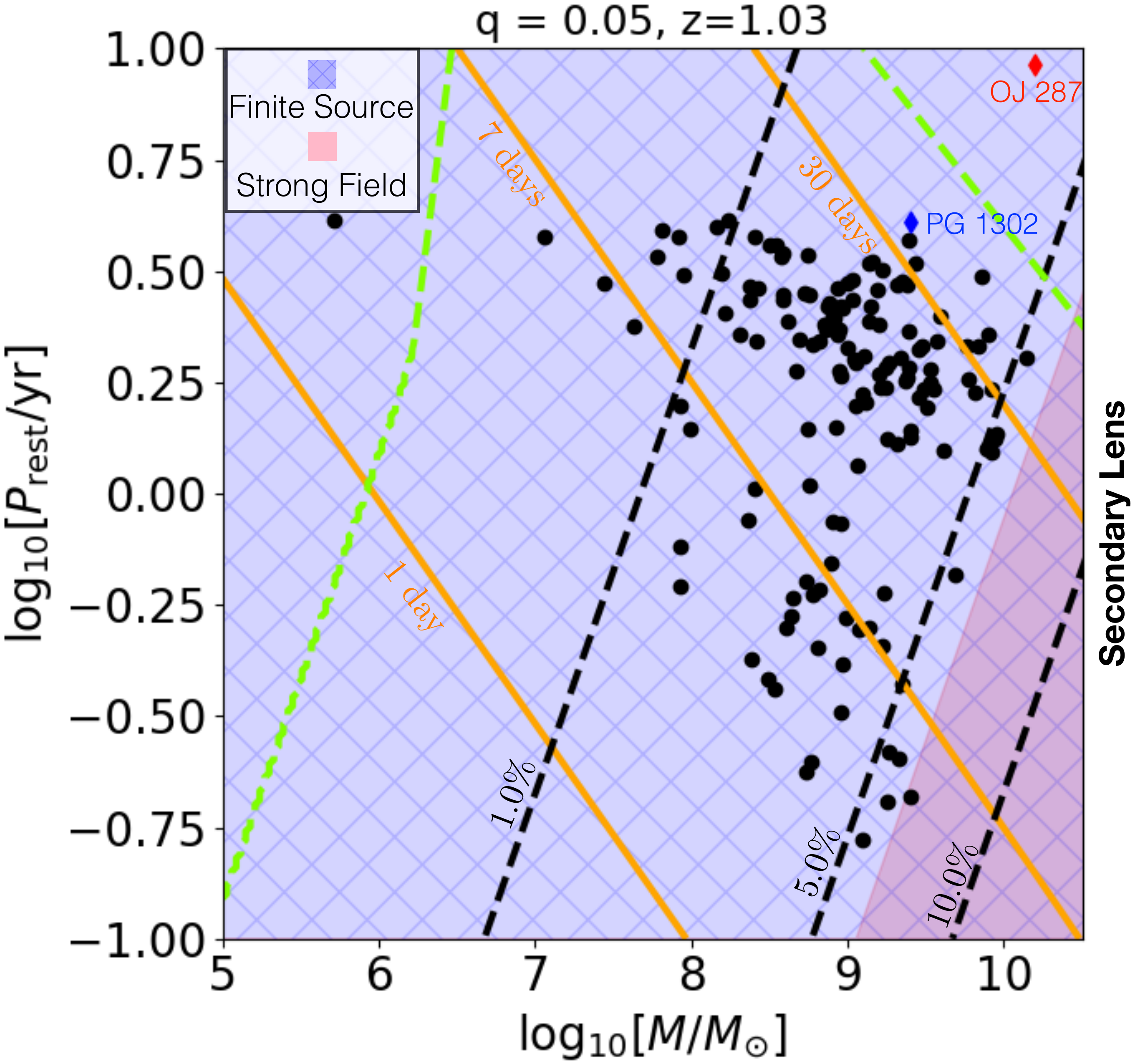} \\
\includegraphics[scale=0.3]{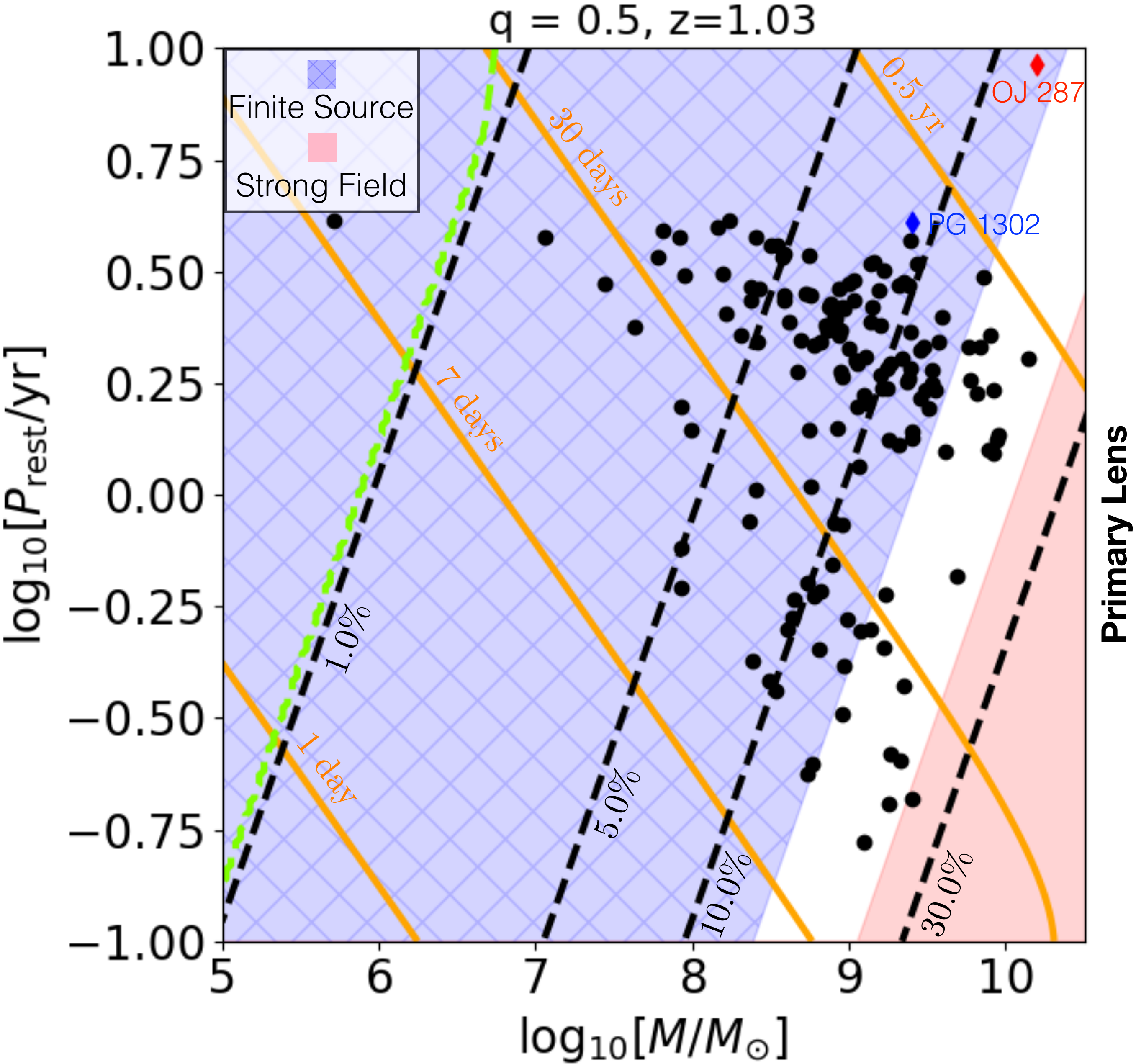} &
\includegraphics[scale=0.3]{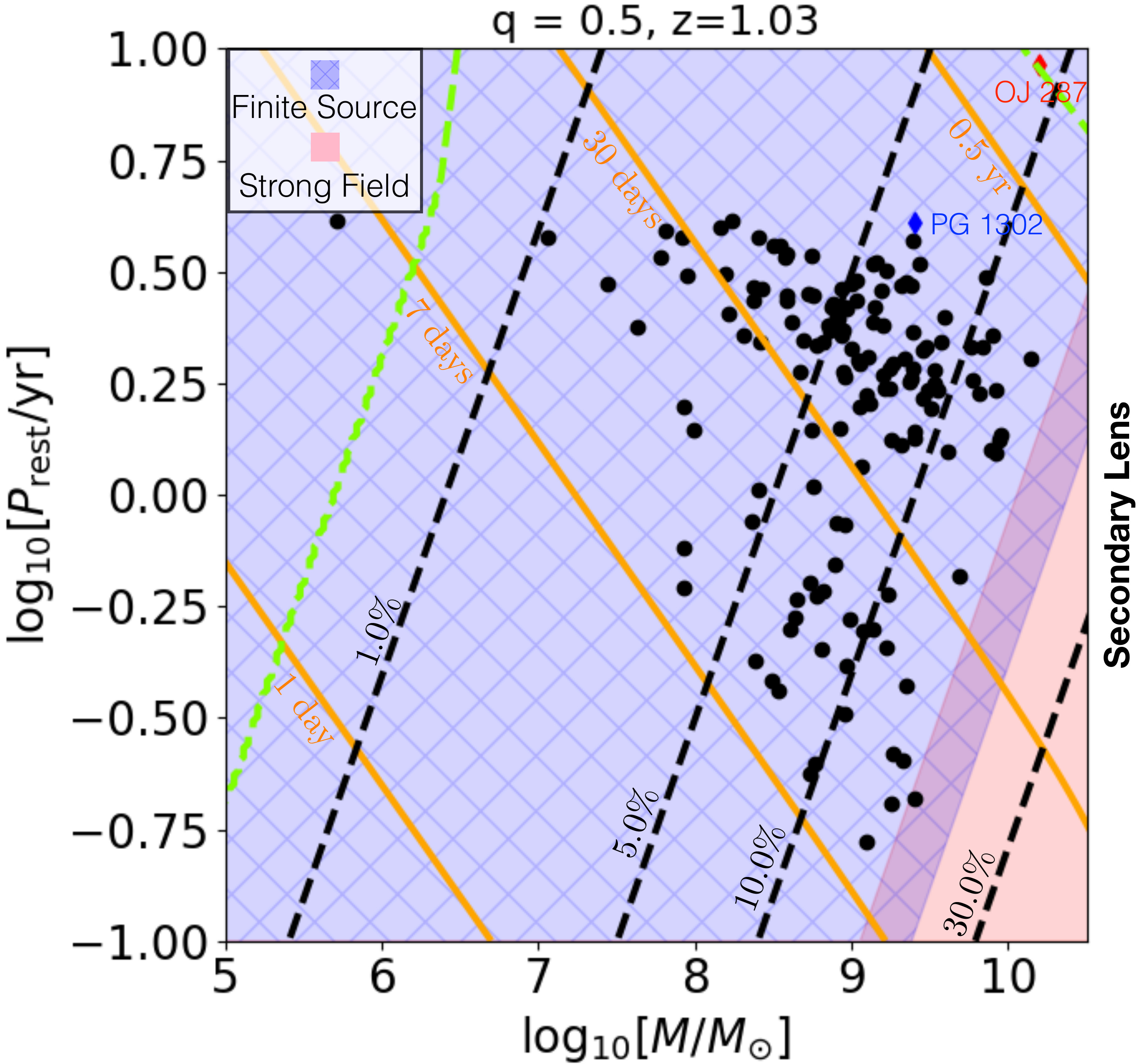} 
\end{array}$
\end{center}
\caption{
Contours in orbital period and binary mass space of the lensing time-scale
(\textcolor{orange}{orange lines}) and probability that the source centre
passes within one Einstein radius of the lens, after one orbital period of
observation (\textbf{black-dashed lines}). The left (right) column is for the
primary (secondary) lensing the secondary (primary) accretion disc. The top
(bottom) row assumes a mass ratio of $q=0.05$ ($q=0.5$). The parameter space
is delineated into three regions where different lensing effects dominate. In
the \textcolor{blue}{blue shaded region}, finite-sized source effects are
important due to the large relative size of the source accretion disc to the
Einstein radius of the lens. The \textcolor{red}{red-shaded region} is where
strong-field gravitational effects become important, and the non-shaded
region is where point-lens and point-source approximations are valid. At
separations larger than those delineated by the \textcolor{green}{green-dashed
line}, the disc may be truncated by self gravity rather than tidal forces.
The black circles are MBHB candidates identified in time domain surveys as
periodically varying quasars \citep{Graham+2015b, Charisi+2016}. All
calculations assume $z=1.03$, the average redshift of the plotted MBHB
candidate sample.
}
\label{Fig:PsandTs}
\end{figure*}

Figure \ref{Fig:PsandTs} plots the probabilities (dashed black lines) and
lensing time-scales (orange lines) for a range of binary masses and periods
amenable to observation through time domain surveys. To compute time-scales and
probabilities we choose mass ratios of $q=0.05$ and $q=0.5$ and a redshift of
$z=1.03$ corresponding to the average of the samples of \cite{Graham+2015b}
and \cite{Charisi+2016}. For reference, the masses and periods of these MBHB
candidates are overplotted as black circles. The delineation of the binary
parameter space into finite-sized source (blue), point-source (white), and
strong-field (red) regimes is discussed in more detail in the next section. We
find that for MBHBs with masses and periods representative of those that can
be identified in time-domain surveys, the probability of a lensing event (Eq.
\ref{Eq:Prob}) is significant, ranging from $\sim 1\%$ chances for the
smallest, $10^5 \Msun - 10^6 \Msun$ MBHBs to 10s of percent for  the highest
mass $\gsim 10^8 \Msun$ binaries.

The magnification of a point source by a point mass is \citep{Pac:1986_Microlens},
\begin{equation}
\mathcal{M}_{PS} = \frac{u^2 + 2 }{u \sqrt{u^2 + 4}},
\label{Eq:PmsMag}
\end{equation}
where $u$ is the angular separation of source and point mass in units of the
Einstein radius. For the orbiting secondary,
\begin{eqnarray}
u_s &=& \mathcal{R}e \left\{ \frac{a}{r^p_E} \sqrt{ \cos^2{\Omega t} +  \sin^2{I} \sin^2{\Omega t} }  \right\}, 
\label{Eq:u0sec}
\end{eqnarray}
where $\mathcal{R}e$ denotes the real part and $r^p_E$ is given by Eq. (\ref{Eq:rsE}).  
In the case that the secondary
lenses the primary, we simply swap $r^p_E$ with $r^s_E$ and $\Omega t$ with
$(\Omega t - \pi$). The total magnification of both lensing events is then
found by inserting $u = \mathcal{R}e\{u_s + u_p\}$ in Eq (\ref{Eq:PmsMag}).
The binary inclination angle written in terms of the projected angle on the sky in
units of Einstein radius is, 
\begin{eqnarray}
\sin{I} &=&  A \left[ \frac{\sqrt{4+A^2}}{2A} - \frac{1}{2}\right]^{1/2}; 
\quad A \equiv N^2_E \frac{2R^p_S}{a},
\end{eqnarray}
where $N_E$ is defined to be the inclination angle measured in units of the
maximum Einstein radius of the primary, $r^{\rm{max}}_E$ (see discussion below
Eq. \ref{Eq:Prob}). Equivalently, $N_E$ is the number of Einstein radii
separating source and lens at closest approach. In the limit of of small $I$,
\begin{equation}
\sin{I} \approx N_E \sqrt{\frac{2R^p_S}{a}}.
\label{Eq:INE}
\end{equation}

\begin{figure*}
\begin{center}$
\begin{array}{c c}
\hspace{-0.2cm}
\includegraphics[scale=0.4]{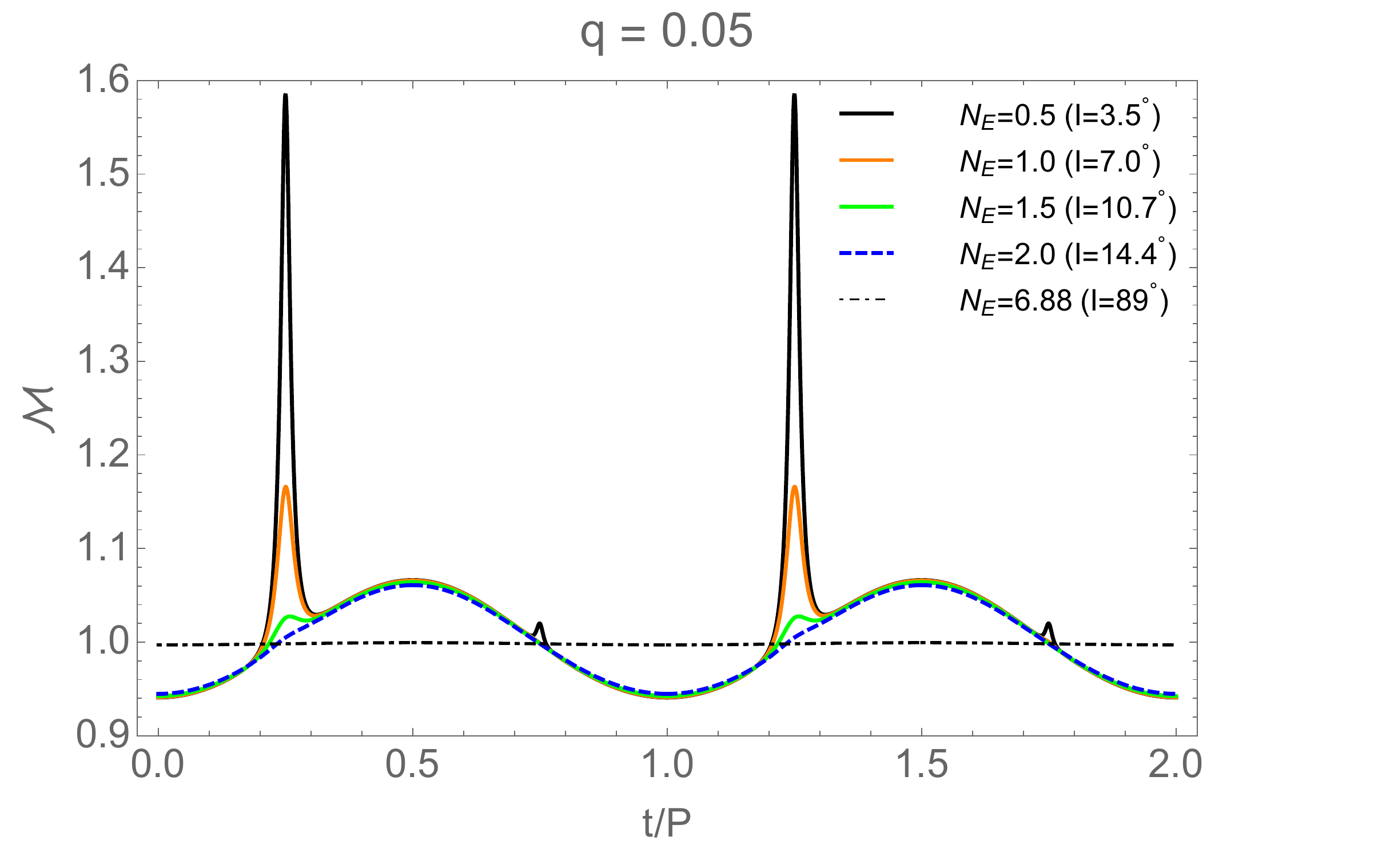} & \hspace{-1.3cm}
\includegraphics[scale=0.4]{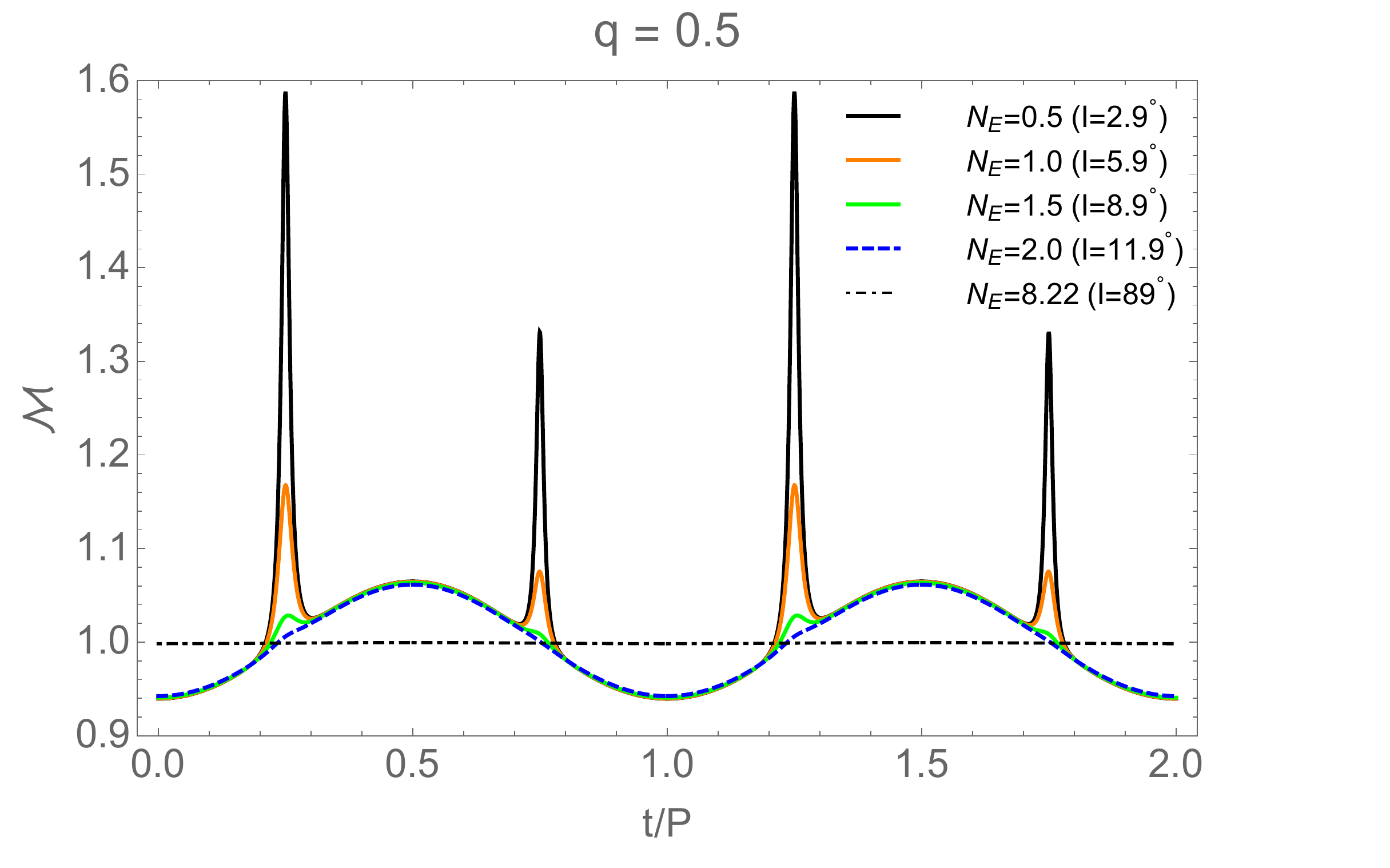} 
\end{array}$
\end{center}
\caption{
Magnification of Doppler boosted and lensed emission from primary and
secondary point-source emission for binary parameters $(P,M)=(4.0 \rm{yr},
10^9 \Msun)$ and for the stated values of the binary orbital inclination,
listed in the legend in terms of $N_E$ defined in Eq. (\ref{Eq:INE}). Both
primary and secondary discs are assumed to emit at the same brightness. The
first lensing flare is due to lensing of the secondary accretion disc by the
primary and the second lensing flare is due to lensing of the primary
accretion disc by the secondary. The inclination of the binary that could
generate the orange light curve has probability given by the black-dashed
lines in Figure \ref{Fig:PsandTs}. The black light curve has a lower
probability of occurring, but would be easier to discover.
}
\label{Fig:PS_LCs}
\end{figure*}

Figure \ref{Fig:PS_LCs} uses Eqs. (\ref{Eq:PmsMag}) and (\ref{Eq:u0sec}) to
compute sample light curves of observed emission from a steadily accreting
secondary in a binary with a total mass of $10^9 \Msun$ and a period of $4$
years. We assume the case of Doppler-boost induced broad-band periodicity: we
allow both primary and secondary BH to emit at the same total brightness and
combine the competing contributions from the relativistic Doppler boost, each
multiplied by the appropriate lensing magnification
\citep[see][]{PG1302Nature:2015b}. In each panel we plot light curves for
different binary inclinations to the line of sight. Note that, because the
total emitted light is split between the two binary components, the maximum
magnification of a lensing flare is smaller than if we assumed all light came
from one component (compare the peaks in Figure \ref{Fig:PS_LCs}, with the
corresponding values in Figure \ref{Fig:Mags_FS}).

In the left panel of Figure \ref{Fig:PS_LCs}, we assume a binary mass ratio
of $q=0.05$, then for $N_E \lsim 1.5$ ($\sim 11^{\circ}$ from edge on) we find
an appreciable lensing flare as the secondary passes behind the primary,
reaching a factor of $\sim1.6$ in magnification when $N_E \rightarrow 0.5$
($\sim 4^{\circ}$ from edge on). In the right panel we consider a larger,
$q=0.5$, mass ratio binary. In this case we see two lensing flares. The first
is similar to the more extreme mass ratio case, corresponding to when the
secondary passes behinds the primary, and a second flare now occurs as the
primary (if it is accreting) passes behind the secondary. In both cases, the
lensing flares disappear as the binary inclination increases, first for the
lensed primary and then for the lensed secondary, leaving only the sinusoidal
Doppler-boost signal (dashed-blue line) which then also disappears as the
binary inclination approaches face on (dot-dashed black line).

We see only a small secondary flare in the more extreme mass ratio case (right
panel of Figure \ref{Fig:PS_LCs}) because, for small $u$, $\mathcal{M}_{PS}$
(Eq. \ref{Eq:PmsMag}) goes as $1/u$ and $u \propto 1/r_E \propto M^{-1/2}$,
then the maximum lensing magnification by the secondary BH (second, smaller
flare) is related to the maximum lensing magnification by the primary BH
(first, larger flare) by $\sqrt{q}$. Hence the binary with $q=0.05$ does not
exhibit a significant secondary flare as this flare is $\sim4.5$ times smaller
than the primary flare. The $q=0.5$ (right panel) binary does exhibit a
secondary flare as its magnification is only $\sim 1.4$ times smaller than the
primary flare.

A key feature of the light curves in Figure \ref{Fig:PS_LCs} is the time-order
of the flares and their magnitudes: a high magnification flare followed by a
lower magnification flare (or no detectable flare at all). For a Doppler-boost
plus lensing light curve this time and magnitude ordering always occurs unless
the primary is emitting a significantly larger fraction of the light in the
observing band. This is because of the following reasons.
\begin{enumerate}
\item{The more massive primary generates a larger magnification of the
secondary accretion disc than in the opposite scenario.}
\item{
Because the secondary is expected to accrete at a higher rate
\citep{Farris:2014} and moves faster in its orbit, it dominates the
sinusoidal Doppler-boost signature. Hence, the sinusoidal part of the light
curve in Figure \ref{Fig:PS_LCs} is increasing after the secondary is lensed,
when the secondary is moving towards the observer.
}
\end{enumerate}
The above two points imply that the secondary passes behind the primary
during the increasing flux portion of the sinusoidal light curve, and hence
the larger flare occurs here. For flares that accompany broad-band periodicity
due to accretion variability, the time-ordering of flares relative to the
light curve minima and maxima is not constrained. This ordering could inform
us how accretion proceeds on to the binary components.

Figure \ref{Fig:PsandTs} displays lensing time-scales and probabilities but does
not show the maximum lensing magnification as a function of binary mass and
period. This is because magnification depends solely on the closest angular
approach of the secondary and primary on the sky ($u$), and hence depends only
on the binary inclination to the line of sight. Writing the inclination in
terms of angular Einstein radii $N_E$ (Eq. \ref{Eq:INE}), the solid black line
in Figure \ref{Fig:Mags_FS} shows the maximum magnification of the secondary
point source as a function of $N_E$. For example, when $N_E=1.0$, $\mathcal{M}
\approx 1.3$, or when $N_E=0.5$, $\mathcal{M} \approx 2.2$ (larger than seen
in Figure \ref{Fig:PS_LCs} because the Doppler calculation assumes that each
binary component contributes equally to the total brightness of the system).
The role of binary mass and period in the point-source case is to gives
physical units (in radians) to the inclination in units of angular Einstein
radii. This is exactly the information portrayed by the probability contours
in Figure \ref{Fig:PsandTs}. A larger binary orbital separation (lager binary
period and smaller mass) means that that a smaller range of binary
inclinations corresponds to the same inclination (and hence maximum
magnification) in units of Einstein radii.

\subsection{Finite Source}
\label{S:FiniteSource}

The emitting, lensed accretion disc can no longer be treated as a point source
if its angular extent is of order the Einstein radius of the lensing BH,
\begin{equation}
\rho(\lambda) \equiv \frac{r_d(\lambda)}{r_E} \approx \frac{1}{\sqrt{2 n_a}}\frac{r_d(\lambda)}{R_S}  \gsim 1 ,
\end{equation}
where again $n_a$ is the binary separation in units of the Schwarzschild radius of
the total binary mass and $r_d(\lambda)$ is the wavelength-dependent size of
the accretion disc.

We model $r_d(\lambda)$ by assuming a steady-state alpha disc where
absorption opacity dominates scattering. Then using the standard steady-state
disc photosphere temperature (see Eq. \ref{Eq:ADTemp} below) for $r \gg 6
GM/c^2$, we solve for the radius where the accretion disc spectrum peaks at
wavelength $\lambda$. Dividing by the Einstein radius of primary and secondary
respectively,
\begin{eqnarray}
\rho^s(\lambda) &\approx& 1.1  \frac{q^{2/3}}{(1+q)^{1/6} } \left( \frac{150}{n_a} \right) \left(\frac{\lambda}{230 \rm{nm}} \right)^{4/3} \left( \frac{\epsilon^{-1} \eta M}{10^9 \Msun}\right)^{-1/3}    \nonumber  \\
\rho^p(\lambda) &=& q^{-7/6} \rho^s(\lambda)
\label{Eq:rhoLambda}
\end{eqnarray}
where again superscripts $s$ and $p$ refer to the binary component being lensed,
$\epsilon$ is the disc accretion rate in units of the Eddington rate, and
$\eta$ is the accretion efficiency factor.

The size of the disc is limited at the inner edge by the innermost stable
circular orbit (ISCO) of the accreting BH. At the outer edge the disc size is
limited by tidal truncation or possibly by gravitational or ionizational
instability. Assuming for now that tidal truncation dominates (which it does
for smaller binary separations, where finite-sized source effects first become
important, and where the probability of lensing is larger), we explore the
importance of treating the source as having a finite-size.

For the secondary disc, the
tidal truncation radius is $r^s_{\out} \approx 0.27 q^{0.3}
a$, and for the primary $r^p_{\out} \approx 0.27 q^{-0.3} a$
\citep{Roedig+Krolik+Miller2014, PapaPringle:1977, PR:InnerDsks:1981}. Then we
may estimate the largest accretion disc sizes at any wavelength as,
\begin{eqnarray}
\rho^s_{\max} &\approx& 1.0 \left( \frac{n_a}{27.5}\right)^{1/2}  q^{ 0.3}  \sqrt{1+q} \nonumber \\
&\approx& 1.0 \left(\frac{M}{10^9 \Msun} \right)^{-1/3}  \left(\frac{P}{0.4 \rm{yr}} \right)^{1/3}  q^{ 0.3}  \sqrt{1+q}, \nonumber \\
\rho^p_{\max} &=& q^{-1.1} \rho^s_{\max} .
\label{Eq:rho_max}
\end{eqnarray}

The peak wavelength of emission at the accretion disc outer edge can be found
by equating $r_{\rm{out}}$ and $r(\lambda)$ used in Eq. (\ref{Eq:rhoLambda})
above,  
\begin{eqnarray} 
\lambda^s_{\rm{out}} &=& 209 \rm{nm}   \left( \frac{M}{ 10^9 \Msun }\right)^{-1/4} \left( \frac{P}{4.0 \rm{yr} }\right)^{1/2} \left(\frac{\eta}{\epsilon}\right)^{1/4}q^{9/40} \sqrt{1+1/q} 
\nonumber \\
\lambda^p_{\rm{out}} &=& q^{1/20} \lambda^s_{\rm{out}}.
\label{Eq:Lam_out}
\end{eqnarray}

The blue shaded regions of Figure \ref{Fig:PsandTs} show where tidal
truncation predicts that finite-sized source effects become important using
Eqs. (\ref{Eq:rho_max}). We see that the finite size of either primary or
secondary accretion disc can become important for fiducial binary masses,
periods, and mass ratios considered here and also at important observing
wavelengths via Eq. (\ref{Eq:Lam_out}). Accretion discs around lower mass, and
longer period binaries are lensed as finite-sized sources unless disc
instability such as self-gravity truncates the disc to a smaller size than the
tidal truncation radius. Also, lensing of the primary accretion disc by the
secondary is more likely to occur in the finite source regime.

Note that $\rho_{\max}$ is an increasing function of binary separation. This
follows because a wider binary allows the discs around each BH to be truncated
at a larger physical distance from the BH that grows linearly with the orbital
separation, whereas the Einstein radius grows with the square root of the
orbital separation. The interpretation of $\rho_{\max}$ as an indicator of the
importance of finite-sized source effects then follows only if the disc is
stable out to the truncation radius. This assumption could break down at
sufficiently large binary separations.

As an estimate of where instabilities could truncate the disc to smaller sizes
than required by tidal truncation, we calculate the mini-disc gravitational
instability radii assuming steady-state alpha-disc solutions (Eqs (12-16) of
\cite{HKM09}). The binary parameters for which the tidal and self-gravity
truncation radii are equal are delineated as a green-dashed line in each
panel of Figure \ref{Fig:PsandTs}. This shows that the smallest and largest
mass and longest period binaries could have mini-discs that are truncated by
gravitational instability, causing them to act as point sources rather than
the finite-sized sources predicted by tidal truncation. We note,
however, that this instability radius is highly uncertain as it assumes a
specific steady-state disc model. We continue our analysis using the more
robust tidal truncation radii, but wish only to point out that self-gravity
truncation could become important for determining the lensing regime for the
binary parameters of interest.

To compute the wavelength-dependent magnification of the secondary's accretion
disc, we take into account the radius and wavelength-dependent flux given by 
steady-state alpha-disc models,
\begin{eqnarray}
F_{\nu}(r) = \pi  B_{\nu}[T(r)],
\end{eqnarray}
where $B_{\nu}$ is the Planck function and $T(r)$ is the radius-dependent 
effective temperature of the disc photosphere, given by balancing viscous
heating with radiative cooling,
\begin{eqnarray}
\sigma T^4 = \frac{3GM \dot{M}}{8 \pi r^3}\left[1 - \left( \frac{r_{\rm{ISCO}} }{r}\right)^{1/2} \right].
\label{Eq:ADTemp}
\end{eqnarray}
Using polar coordinates $(u,v)$ centred on the lens, and $(r,\phi)$ centred
on the source, the magnification, in a given observing band is
\begin{equation}
\mathcal{M}^{FS}_{\nu} = \frac{ \int^{2 \pi}_0 \int^{\infty}_{0}{ F_{\nu}(u', v') \mathcal{M}_{PS}(u') u' \ du' }    dv' }
{ \int^{2 \pi}_0 \int^{\infty}_{0}{ F_{\nu}(u', v') u' \ du' }    dv' } ,
\label{Eq:MAgFS}
\end{equation}
where $\mathcal{M}_{PS}$ is the point-source magnification, Eq.
(\ref{Eq:PmsMag}). To write $F_{\nu}$ in terms of lens-centred
coordinates $(u,v)$ we use that,
\begin{eqnarray}
r^2_* &=& \left( u^2_0 + u^2 -2 u_0 u \cos{\left(v-v_0\right)} \right) r^2_E, \nonumber \\ 
r &=& r_* \sqrt{   \cos^2{\phi}  +  \frac{ \sin^2{\phi} }{\cos^2{ \left( \pi/2 - J \right)}   }  }, \\ 
\sin{\phi} &\equiv& \frac{u\sin{\nu} - u_0\sin{\nu_0}}{ \sqrt{  (u\sin{\nu} - u_0\sin{\nu_0})^2 + (u\cos{\nu} - u_0\cos{v_0})^2 } },
\nonumber
\end{eqnarray}
where $(u_0, v_0)$ is the position of the secondary in lens-centred, polar
coordinates: $u_0$ is given by Eq. (\ref{Eq:u0sec}) in units of Einstein radii
and $v_0 = \tan^{-1}\left[ \sin{I} \tan{\Omega t}\right]$; $J$ is the
inclination of the source disc to the line of sight (see Figure
\ref{Fig:Schem}), and $\phi$ is the polar coordinate in the disc-centred
frame. To evaluate Eq. (\ref{Eq:MAgFS}), we set $F_{\nu}=0$ when $r$ is less
than the ISCO radius or when $r$ is greater than the truncation radius discussed
above.

\subsubsection{Wavelength-dependent effects}

\begin{figure*}
\begin{center}$
\begin{array}{c c c}
\includegraphics[scale=0.3]{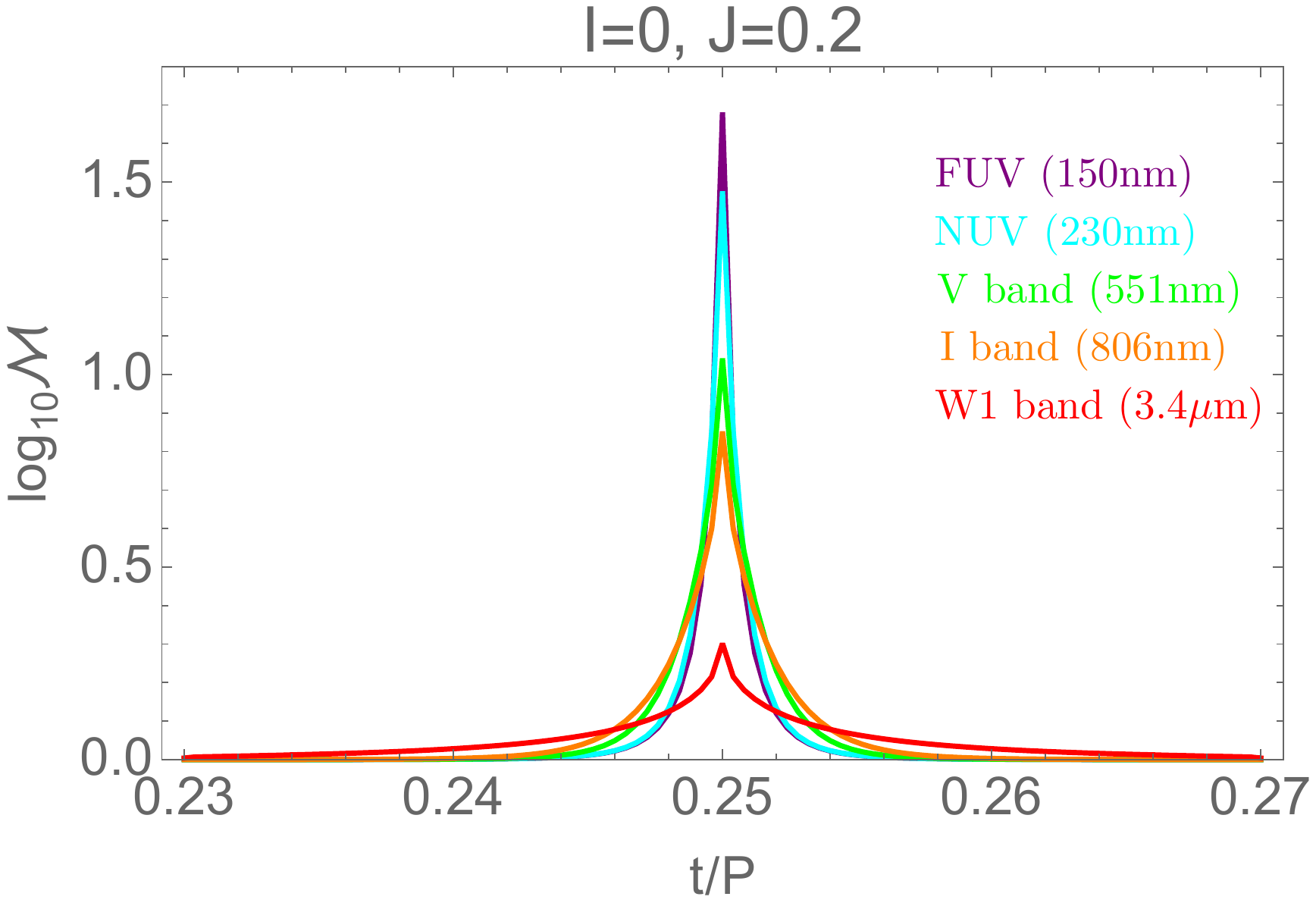} &
\includegraphics[scale=0.3]{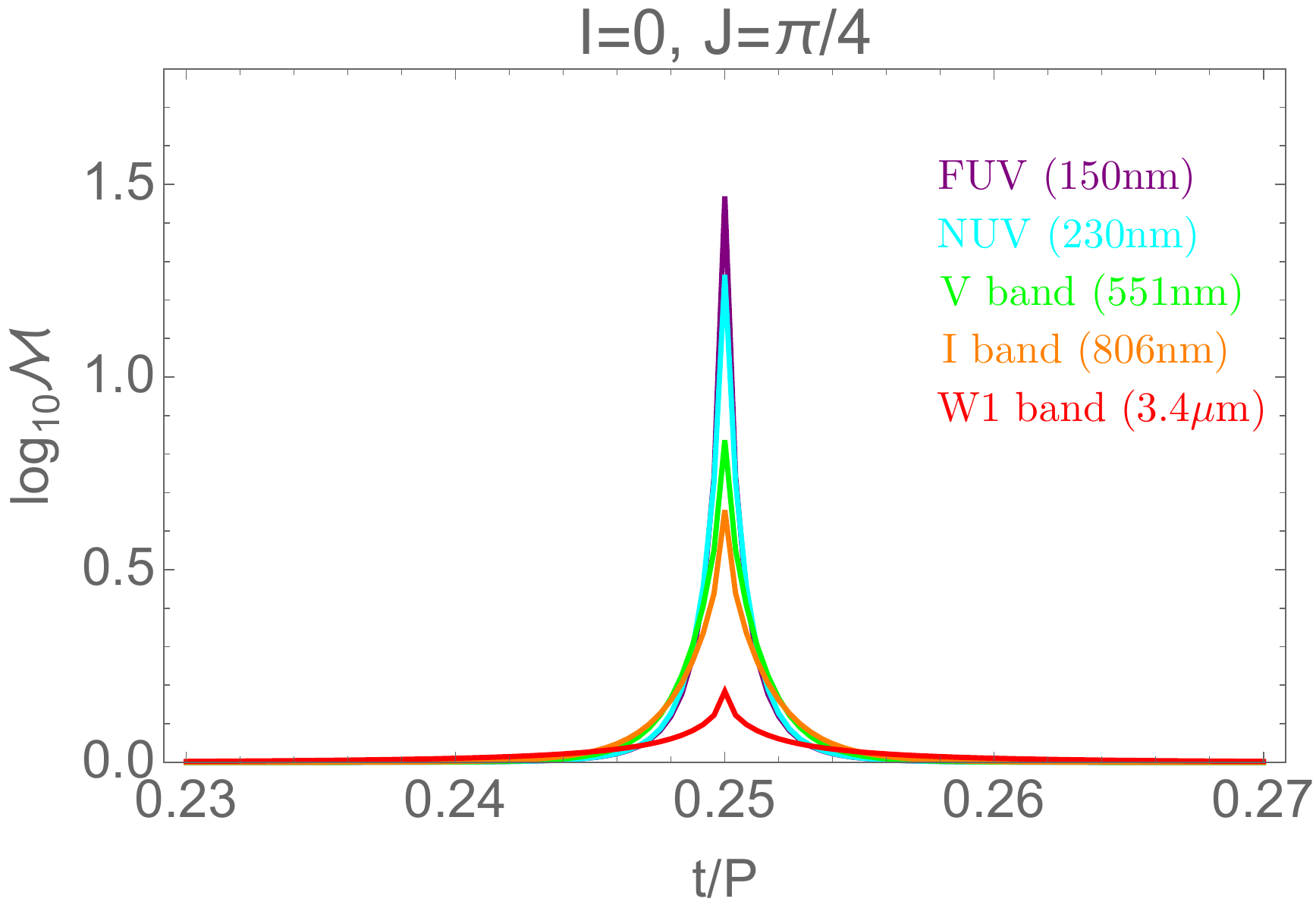} &
\includegraphics[scale=0.3]{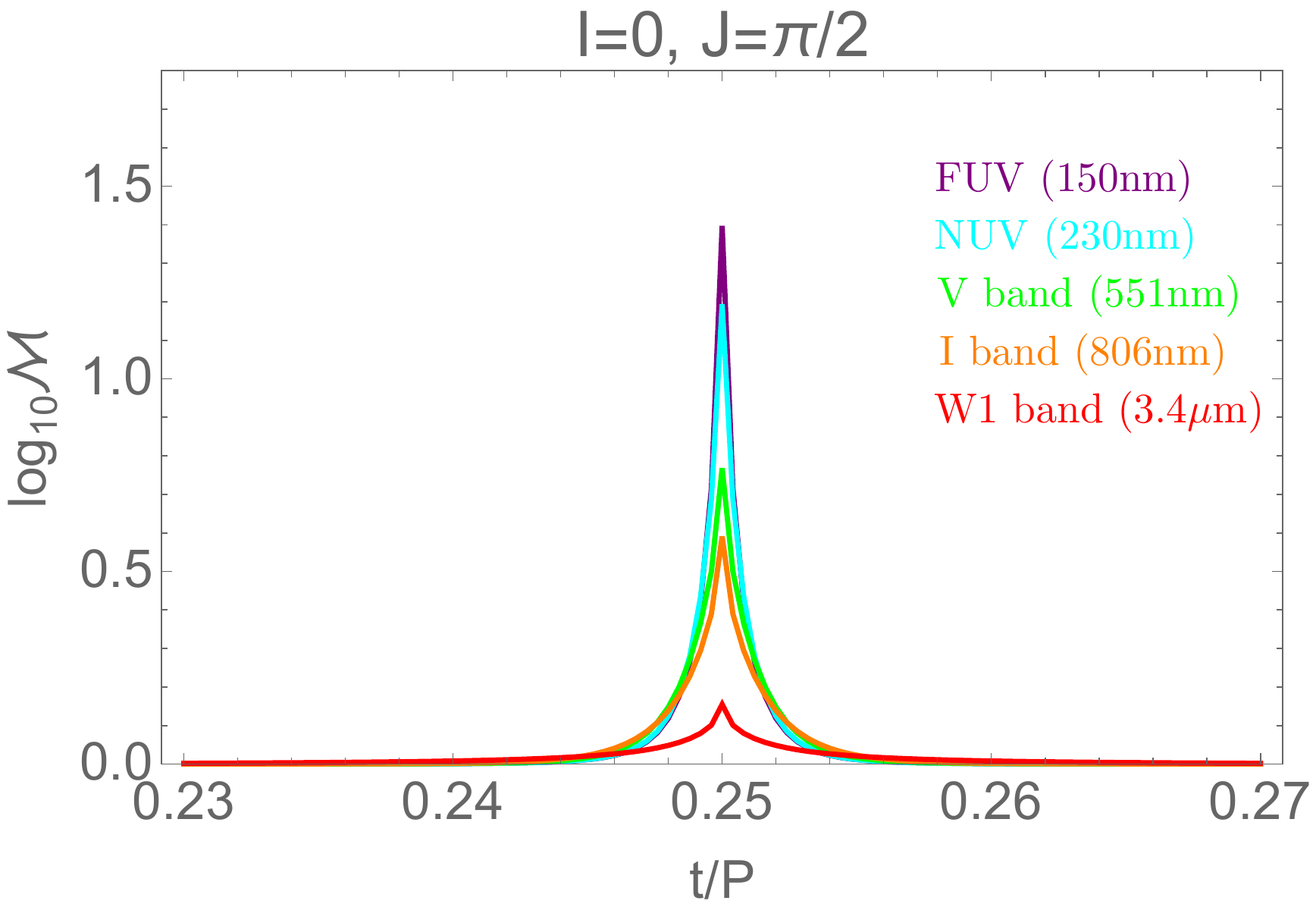} \\
\includegraphics[scale=0.3]{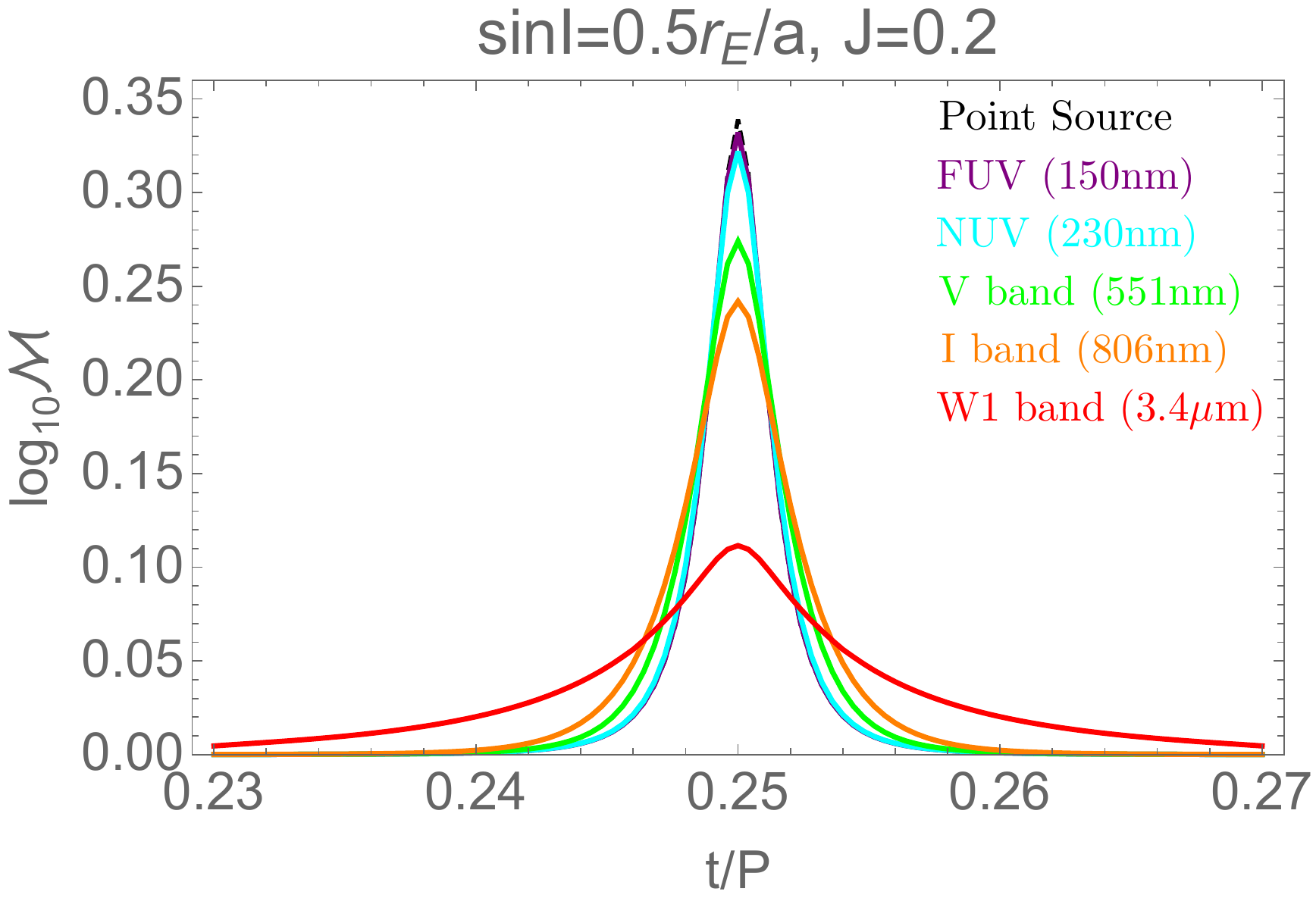} &
\includegraphics[scale=0.3]{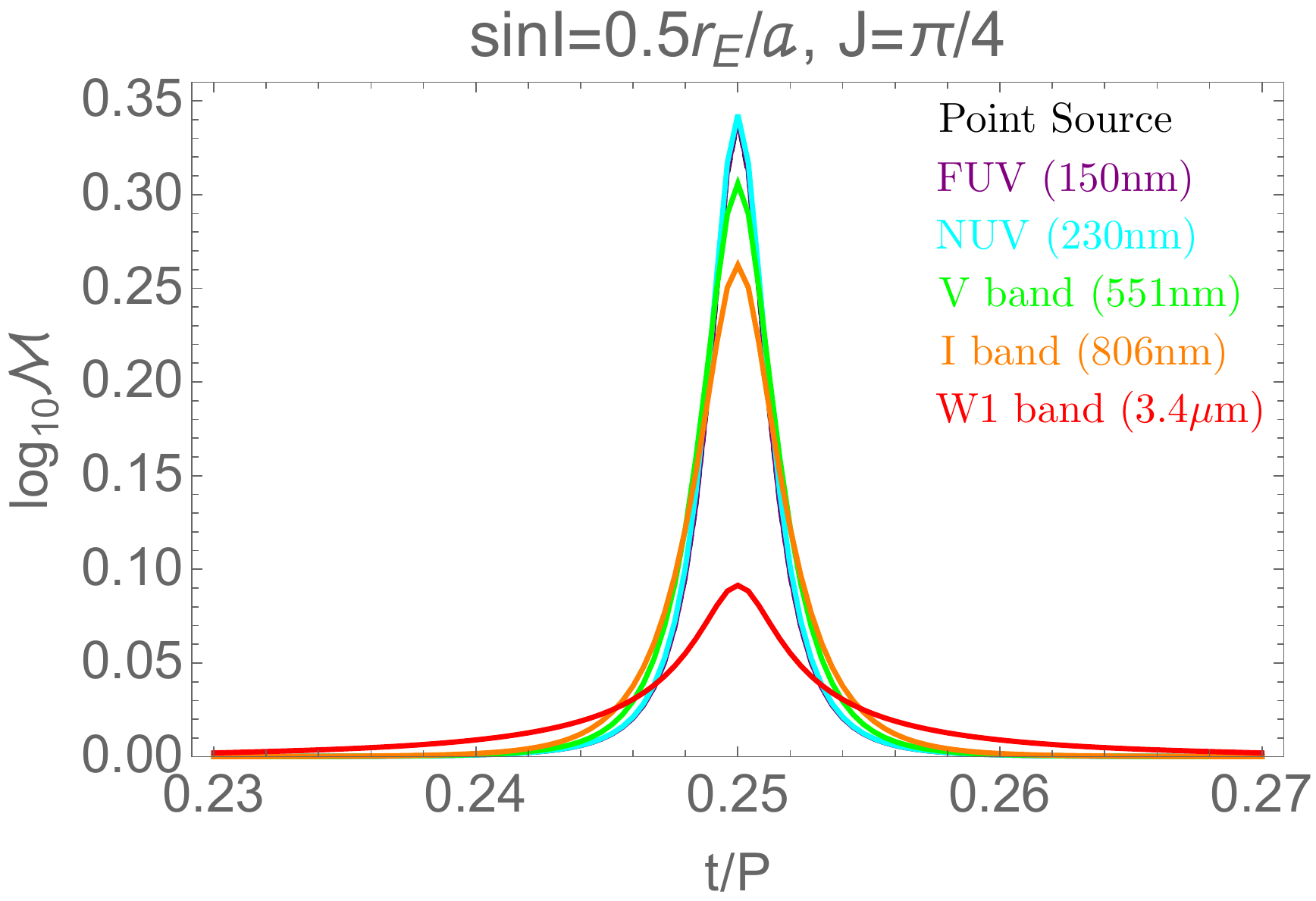} &
\includegraphics[scale=0.3]{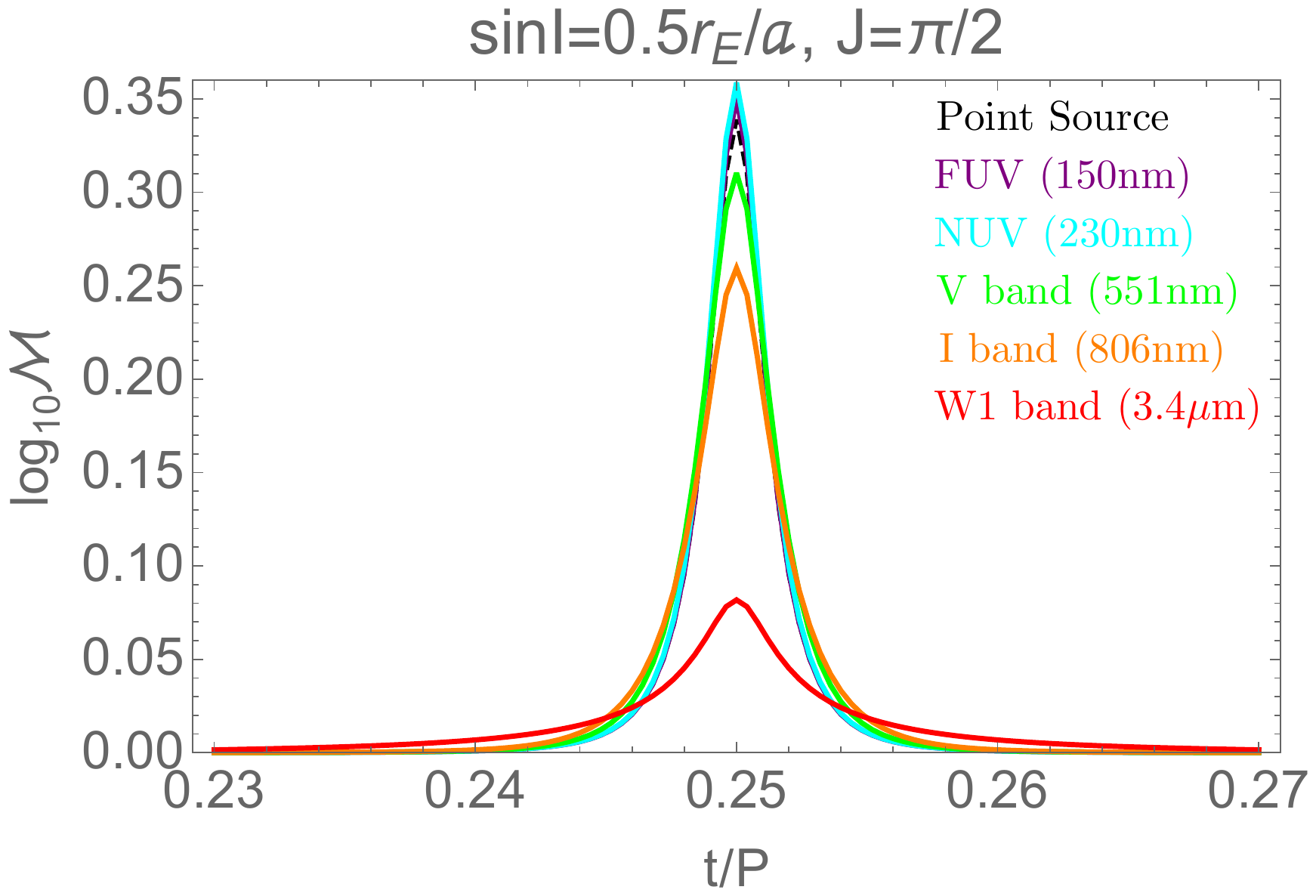}
\end{array}$
\end{center}
\caption{
Zoomed-in light curves for multi-wavelength observations of a lensed, 
finite-sized accretion disc around the secondary BH. Here we fix the binary to have
total mass $10^6 \Msun$, mass ratio of $q=0.1$, and an orbital period of 4
years. This sets the plotted time frame to be approximately two months. Each column,
from left to right, assumes a different value of the secondary accretion disc
inclination to the line of sight, from nearly edge on ($J=0.2$) to face on
($J=\pi/2$). The top row assumes an edge-on binary orbital inclination ($I=0$,
$N_E=0$), and the bottom row assumes a binary orbit inclined such that the
projected sky separation of the BHs at closest approach is half of an Einstein
radius ($\sin{I} \approx 0.5 r_E/a$, $N_E\approx0.5$). For the binary
considered here this corresponds to an $\sim 0.7^{\circ}$ orbital inclination
to the line of sight. We assume that the source accretion disc emits the
spectrum of a standard steady-state accretion disc.
}
\label{Fig:FS_LCs}
\end{figure*}

Figure \ref{Fig:FS_LCs} plots the wavelength-dependent, finite-sized source
lensing light curves, focusing on the flare produced by lensing of the
secondary accretion disc. The top row is for an edge-on binary inclination
while the bottom inclines the binary to $\sim 0.7^{\circ}$ from edge on,
corresponding to a minimum sky separation of $0.5 r^p_E$ ($N_E=0.5$). Both
rows vary the inclination of the accretion disc source to the line of sight
from nearly edge on ($J=0.2$) to face on ($J=\pi/2$).

In each of the panels we have assumed a binary with total mass $10^6 \Msun$,
mass ratio of $q=0.1$, and an orbital period of 4 years. According to Eq.
(\ref{Eq:Lam_out}), this places the peak wavelength of emission at the tidal
truncation radius of the secondary at $\lambda_{\rm{out}} = 2.3\mu$m, in the
near-infrared (IR). The tidal truncation radius of the secondary accretion disc is
$r^s_{\rm{out}} \sim 0.14 a$ corresponding to $\rho^s_{\max} = 11.3$, hence
finite-sized source effects should be prominent in the optical and UV
wavelengths, approaching the point-source case for far-UV and X-ray
wavelengths.


In Figure \ref{Fig:FS_LCs} we plot the simulated lensing flare in far
ultraviolet (FUV; 150 nm) to red optical (I band; 806 nm) and IR (W1 band;
$3.4 \mu$m). Because the disc is hotter at its centre, and hence emits shorter
wavelength radiation there, the FUV represents the smaller inner region of the
secondary disc while the optical/IR represents the larger outer portion of the
disc. The result, seen in Figure \ref{Fig:FS_LCs}, is that the optical/IR lensing
is spread out, beginning earlier, ending later, and reaching a lower peak
magnification than the flares at shorter wavelengths. The shorter wavelength,
NUV and FUV flares approach the shape and peak magnification predicted by the
point-source case.


The finite-sized source lensing depends not only on the size of the source, but
also on the shape. The inclination of the disc to the line of sight changes the
shape of the emitting region from a large face-on disc at $J=\pi/2$ to a more
narrow ellipse constricted in the direction perpendicular to motion across the
sky (however, see \S \ref{S:Extensions}). The result is that more edge-on discs
approach more closely the  point-source case because the elongated edge-on
discs are less smeared out over the $u=0$ caustic. This can be seen in Figure
\ref{Fig:FS_LCs} where, from left to right, the disc is inclined from nearly
edge on to face on and the redder wavelength light curves decrease in peak
magnitude. The width of the lensing curve does not change greatly because we
only allow the disc to be inclined to the line of sight. Future parameter
studies could allow a more general parametrization of the accretion disc
shape: this could result in a larger range of possible flare durations.


We investigate the effect of changing binary inclination angle by comparing
the top and bottom rows of Figure \ref{Fig:FS_LCs}. Just as for the point-source
case depicted in Figure \ref{Fig:PS_LCs}, increasing the binary inclination
away from the line of sight decreases the peak magnitude of amplification. In
the finite-sized source case, increasing the binary inclination also rounds
the light curve from a cuspy spike at edge-on inclination, caused by a direct
crossing of the source through the $u=0$ caustic, to a rounder flare when the
source passes just above or below the caustic. In the bottom row, for which
the binary is inclined from edge on, we have drawn the
point-source case as a dashed black line. We do not draw the point-source
light curve for the edge-on binary because it results in an infinite
magnification. This infinite magnification is smeared out for the finite-sized
source cases plotted in Figure \ref{Fig:FS_LCs}.

For a binary inclined to the line of sight ($I\neq0$), Figure \ref{Fig:FS_LCs}
shows that the magnification of the shorter wavelength emission does not
change appreciably with $J$, and is similar to the point-source case. The
longer wavelength emission, however, is magnified over a shorter time-scale as
$J$  is increased towards face on. This shows that binary orbital inclination
affects different wavelength emission differently in the finite-sized source
case. Generally, if the extended source, at a given wavelength, has components
that intersect the lensing caustic, these wavelength components are affected
by finite-size source effects; it is the shape of the emitting region on the
sky, and hence $J$, that determines whether or not different regions of the
disc cross the $u=0$ caustic.

\begin{figure*}
\begin{center}$
\begin{array}{c c c}
\includegraphics[scale=0.3]{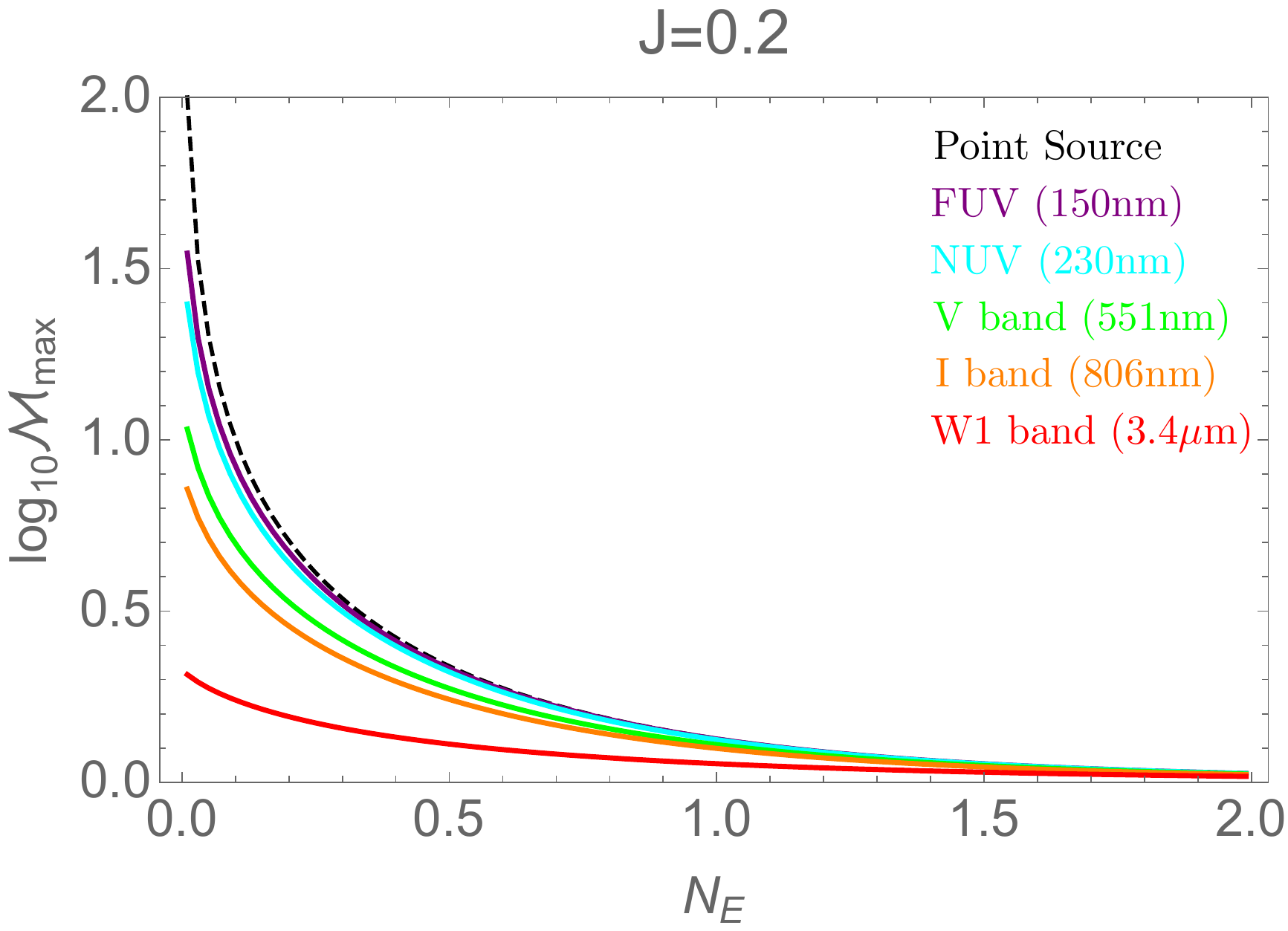}  &
\includegraphics[scale=0.3]{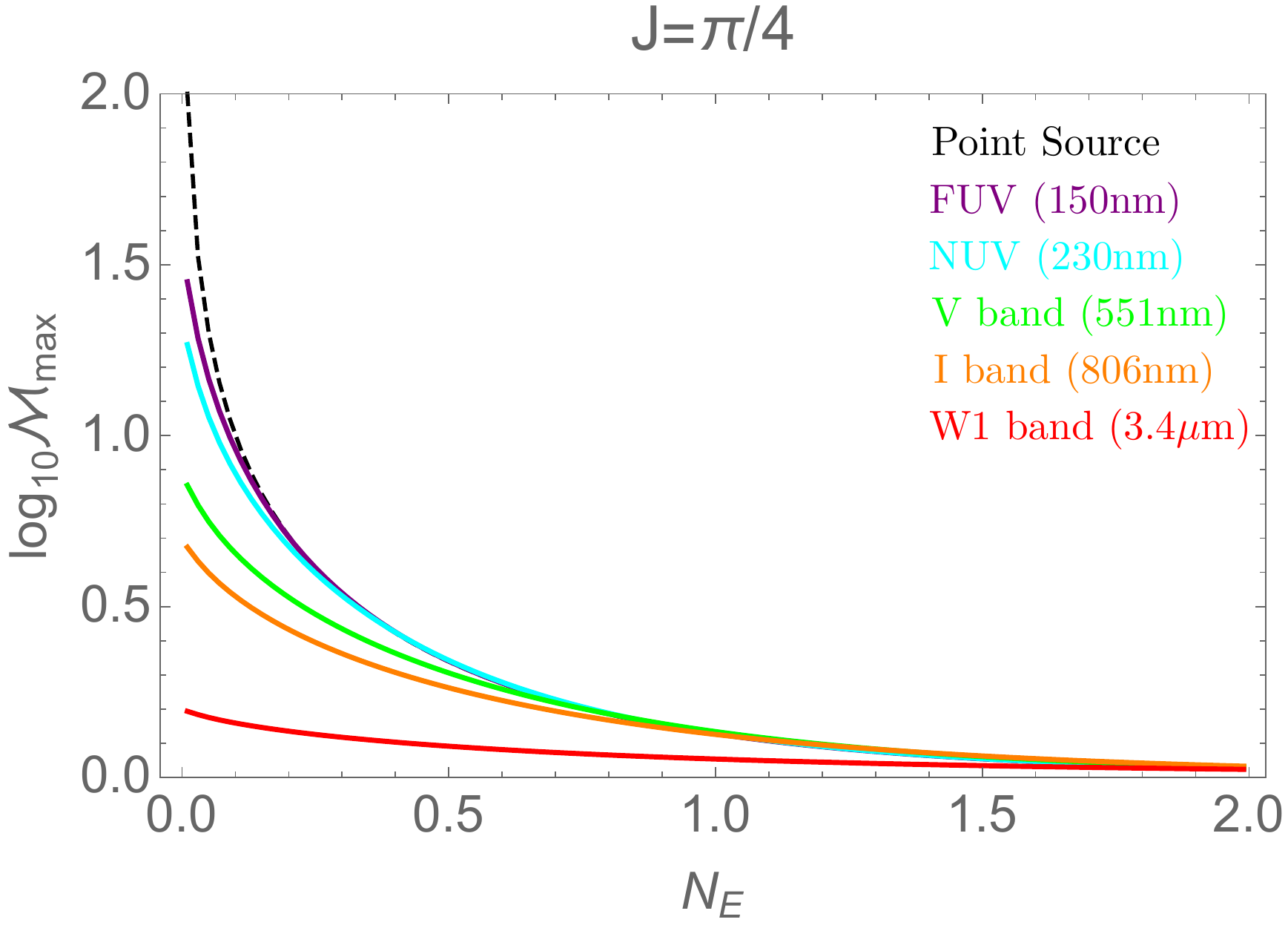}  &
\includegraphics[scale=0.3]{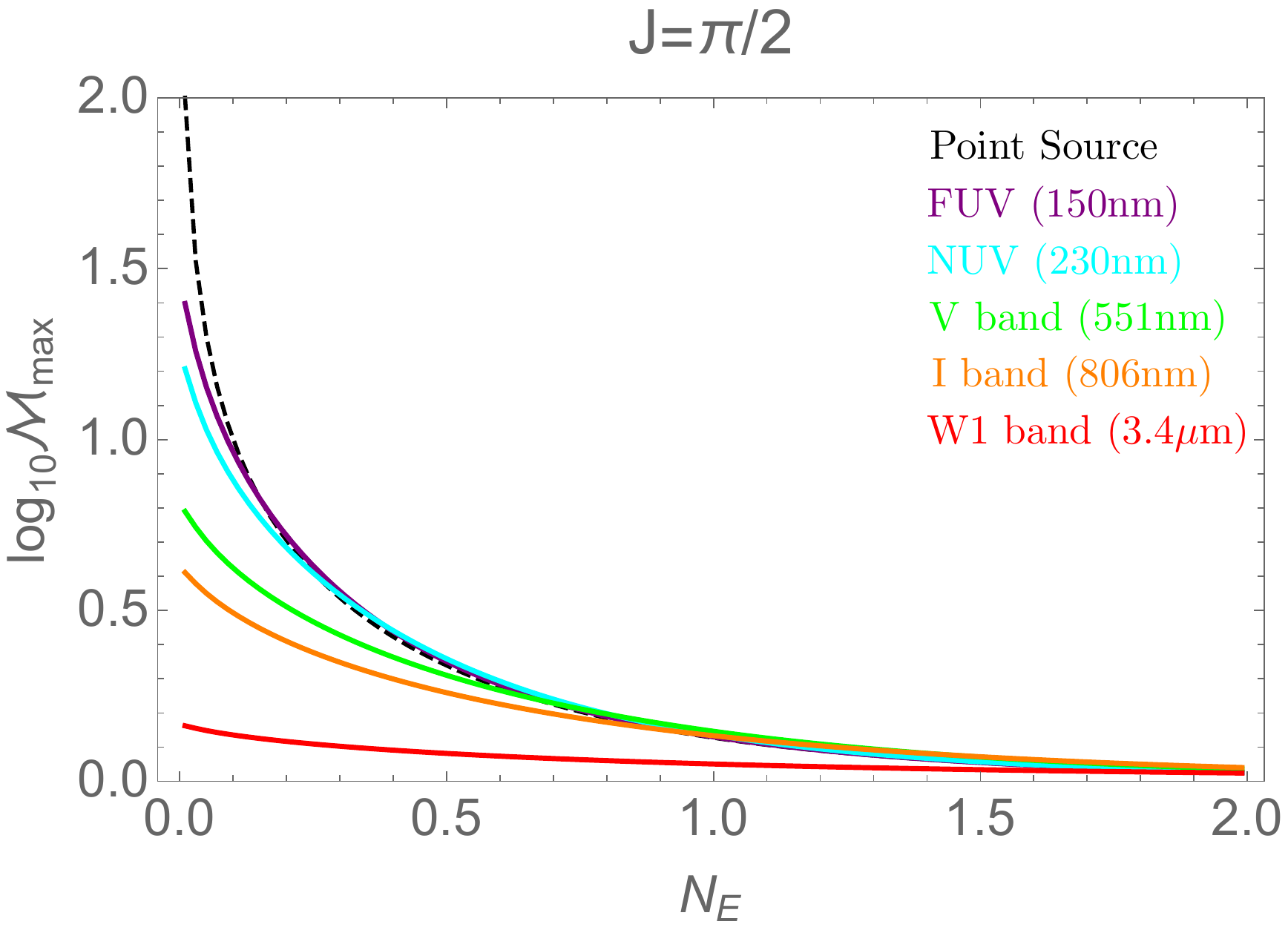} 
\end{array}$
\end{center}
\caption{
Magnification of the secondary emission, at closest approach to the lens,
plotted against the inclination of the binary to the line of sight in units of
the number of Einstein radii that make up the angular separation of source and
lens at closest approach, $N_E$. The finite size of the source is set by the
observed wavelength. Shorter wavelength radiation comes from a smaller inner
region of the disc and so approaches the result of the point-source
calculation (dashed-black line). The binary mass and period set the inclination
angle in radians (see Eq. (\ref{Eq:INE})). From left to right we  tilt the
accretion disc from a nearly edge-on ($J=0.2$) to a face-on inclination
($J=\pi/2$).
}
\label{Fig:Mags_FS}
\end{figure*}

Figure \ref{Fig:Mags_FS} shows how the maximum magnification in the finite-sized
source case compares to that of the point-source case at different wavelengths,
as a function of binary inclination angle. As expected, the finite-sized source and
point-source cases are identical for large binary inclination, but quickly
diverge once the binary inclination places the source within one Einstein
radius of the lens. As the binary inclination approaches edge on, the 
point-source case approaches infinite magnification while progressively longer
wavelength observations exhibit lower peak magnification as explained
above.

Finally, we note that Figures \ref{Fig:Mags_FS} and \ref{Fig:FS_LCs} only
consider finite-sized source effects for one set of binary mass, mass ratio,
and orbital period. The primary difference in changing binary parameters is
the mitigation of finite-sized source effects for more closely separated MBHBs
(and widely separated MBHBs if self-gravity truncates discs around each BH).
This is apparent from Eq. (\ref{Eq:rho_max}) and Figure \ref{Fig:PsandTs}.
Changing the binary parameters will also relabel the important wavelengths for
finite-sized source effects determined through Eq. (\ref{Eq:Lam_out}).

\subsection{Strong-field effects}
\label{S:Strong-field effects}
Thus far, we have treated the lens as a point mass in the weak-field gravity
limit. We briefly address the validity of this assumption.

Strong-field effects can generate relativistic images not captured in this
analysis, however these images are highly demagnified compared to those
considered here in the weak-field limit. For the MBHB system, the ratio of
weak-field and relativistic magnifications for a Schwarzschild BH is
$\mathcal{M}_{wf}/\mathcal{M}_{rel} \approx 83 (a/R_S)^{3/2}$, where $R_S$ is
the Schwarzschild radius of the lens \citep{Bozza+2001}.

In the strong-field limit, one must also take into account gravitationally
induced time delays that could shift the arrival time and shape of the lensing
signal. For the binary periods considered here, the largest effect, the
Shapiro delay \citep{Shapiro:1964, MeironKL:2017}, is small compared to the
lensing time-scale computed in Eq.
(\ref{Eq:TSL}),
\begin{equation}
\frac{ \delta t_{\rm{Shap}} }{ \mathcal{T}_{s} } \approx 0.006 \left( \frac{M}{10^8 \Msun} \right)^{2/3} \left( \frac{P}{1 \rm{yr} } \right)^{-2/3},
\end{equation}
where we have used a minimum inclination corresponding to the source aligned
with the event horizon radius of the lensing primary. The Shapiro delay
becomes an $\sim10\%$ effect for the relativistic systems near merger
studied in \citep{Schnittman+2017, Haiman:2017}.

Hence, at closer binary orbital separations ($n_a \lsim 10$), future work
must carry out relativistic ray tracing in the binary spacetime
\citep[\eg,][]{Schnittman+2017}. For the purposes of this study, however, the
point mass lens is a good approximation.

\section{Discussion and Conclusion}
\label{S:Discussion}

\subsection{Lensing as a unique signature of MBHBs}

A lensing signature is required by relativity alone, provided the binary is
accreting and at a low orbital inclination to the line of sight. Furthermore,
a self-lensing flare can be distinctly distinguished from intrinsic AGN
flaring and variability by a few unique features. In the point-source case, at
binary separations less than $\sim 100 R_s$ (for the secondary and $q=0.1$),
the amplification caused by lensing is periodic and achromatic, identical in
all wavelengths. Wavelength in-dependent periodic flaring would be a smoking-
gun signature of self-lensing in the point-source regime. Practically, because
intrinsic emission from the lensed source is likely not identical at all
wavelengths, a sufficient understanding of the unlensed light curve must be
modelled from data surrounding the flare in order to confirm a wavelength
invariant amplification.

To elaborate further, consider a representative MBHB candidate from Figure
\ref{Fig:PsandTs}. For such a system, the probability of the secondary passing
within one Einstein radius of the primary is $\sim10\%$. The maximum
magnification in the point-source case is $1.32$, corresponding to a $32 \%$
variability amplitude. Comparatively, the average amplitude of variability in
the combined sample of MBHB candidates from \cite{Graham+2015b} and
\cite{Charisi+2016} is $16\% \pm 8 \%$, smaller than, but comparable to, the
amplification predicted here. This is encouraging for detectability of such
lensing flares. However, the amplification of $1.32$ assumes that all of the
light is emitted by the lensed source. This is likely not the case. As we
discussed at the end of \S\ref{S:Point Source}, this magnification factor
drops to $\sim 1.16$ when it is assumed that the secondary and the primary
contribute equally to the total emission (Figure \ref{Fig:PS_LCs}). If an
equal amount of light is also being emitted by the circumbinary disc, then
this total magnification drops to $1.11$. Furthermore, the relative
contribution from each light emitting component at different wavelengths
causes the lensing signature to have a different peak magnification at these
different wavelengths, even in the point-source case. While this complicates
identification of a lensing signature, it acts as a probe of the spectral
energy distributions of the light emitting components of the MBHB system.

We note further, that we have only discussed the most probable
strongly lensed events, where the source passes within an Einstein radius of
the lens at closest approach (or half of this in Figures \ref{Fig:PS_LCs} and
\ref{Fig:FS_LCs}). Hence, lower probability events generate higher
magnification flares than those plotted in Figure \ref{Fig:PS_LCs} and could
be easier to identify observationally. Generally, there is a trade off between
ubiquity and ease of identification. 

In the case of a finite-sized source, the flare is no longer intrinsically
achromatic in peak magnitude, but does, barring any reprocessing or light-echo
effects, reach peak magnitude at the same time and share similar full width at
zero maxima for all wavelengths (The pulse shapes in Figure \ref{Fig:FS_LCs}
all share nearly identical pulse times from turn on to turn off). If detection
of the orbital Doppler-boost is not possible, this flare-timing coincidence
plus the simultaneous modelling of the accretion disc emission profile and the
wavelength-dependent lensing-flare shape could be carried out in order to
differentiate the lensing event from other AGN flaring activity. If,
for example, such intrinsic AGN flares are generated by a propagating
disturbances in the accretion flow, they will exhibit wavelength-dependent
time lags, not seen in a lensing flare.

As suggested by Figure (\ref{Fig:PS_LCs}), if the orbital inclination of the
binary is small enough to detect a lensing flare, then there will also be
periodic modulation of the light curve due to the relativistic Doppler boost.
If the periodic variation from the relativistic Doppler boost is detected,
more constraints can be placed on the system. In this case, a known mass and
spectral index, in addition to the inferred binary period, yield the Doppler
magnification up to the unknown binary inclination and mass ratio
\citep[\eg,][]{PG1302Nature:2015b}. Then the lensing flares can be identified
by their specific position at the average flux locations of the Doppler
modulation (\eg, Figure \ref{Fig:PS_LCs}) and further constrain the binary
mass ratio, inclination, and the projected source shape (parametrized by $J$
here). A detection of a flare with appropriate properties (width,
achromaticity) at the average flux location of the Doppler sinusoid could even
allow identification of such MBHB self-lensing systems without the need for
multiple cycles, lessening the period requirements discussed in
\S\ref{S:System scales}.

If lensing events from both primary and secondary can be detected along with
the Doppler-boost modulation, then the relative brightness of primary and
secondary accretion discs can be discerned in the given observing band. The BH
that dominates the orbital Doppler modulation (usually the faster moving,
brighter secondary) is lensed during the rising flux portion of the Doppler
light curve and the sub-dominate Doppler contributor is lensed during the
falling flux portion. Whether the dominant or sub-dominant emission is
generated by the secondary or primary can be discerned from the relative
widths and magnifications of each lensing flare, smaller by a factor of
$\sqrt{q}$ for lensing by the secondary.

Eqs. (\ref{Eq:rho_max}) show that for both BH discs to enter the finite source
regime the binary separation must be,
\begin{equation}
a_{\rm{fs}} \gsim 27.5 R_S q^{-0.6} (1+q)^{-1}.
\label{Eq:afs}
\end{equation}
For both to be in the point-source regime the binary separation must satisfy,
\begin{equation}
a_{\rm{ps}} \lsim 27.5 R_S q^{1.6} (1+q)^{-1} .
\label{Eq:aps}
\end{equation}
Hence, for a mass ratio of $q=0.1$,
the binary must have separation above $\sim 100 R_S$ for both primary and
secondary discs to act as finite-sized sources, or separations below $0.6$
Schwarzschild radii for both to act as point sources.

For binary separations $\sim 100 R_S$, the
secondary accretion disc is effectively a point source when lensed by the
primary while the primary accretion disc must be treated as a finite-sized
source when lensed by the secondary. This scenario could prove useful for
identifying the lensing system with the more highly magnified and achromatic
lensing of the secondary disc and then using the finite source lensing of the
primary disc to learn about the accretion flow on to the primary. 

Because $a_{ps}$ is small enough that the primary and secondary discs may be
truncated behind the ISCO, we conclude that the case where both accretion
discs act as a point source is only achieved for near-unity mass ratios, very
close to merger.

\subsection{The MBHB population}

At the time of writing there are $\sim 150$ quasars identified as MBHB
candidates by optical periodicity \citep{LehtoValtonen1996, Graham+2015a,
Graham+2015b, Liu:7RsMBHB:2015, Zheng+2016, Charisi+2016, LiWang:2016,
Dorn-W+2017, Li:Ark120:2017}. Some of these have corroborating UV
\citep{PG1302Nature:2015b}, IR \citep{Jun:2015, DZ:2017}, and radio
\citep{Pursimo+2000}) time series data as well. Figure \ref{Fig:PsandTs} plots
these candidates in period vs. total binary mass space. According then to
Figure \ref{Fig:PsandTs}, at least $\sim 7$ ($5\%$) of these candidates should
exhibit self-lensing flares. While we have not carried out a rigorous search
for such flares in the known candidate light curves, we note that, by eye,
none exhibit the expected behaviour. We do not claim any discrepancy.
However, we discuss why one might expect to find fewer lensing candidates in
the existing sample than predicted by our calculations. We further discuss the
implications of this for MBHB candidates and future searches.

First of all, the MBHB candidates found by \cite{Graham+2015b} and
\cite{Charisi+2016} could be in tension with upper limits on the gravitational
wave background (GWB) measured by the pulsar timing arrays
\citep[PTAs;][]{Sesana+2017}. The present version of the work by
\cite{Sesana+2017}, however, overestimates the mass of the MBHB candidates by
a factor of $\sim4$ (Alberto Sesana and Zolt\'an Haiman; private
communication). This mass overestimate may remove any tension with the GWB
entirely. Any remaining discrepancy between the GWB and the known MBHB
candidates, however, could mean that some of the candidates are manifestations
of red-noise rather than true periodicity or resultant from non MBHB induced
periodicity due to, for example, scaled up versions of the quasi-periodic
oscillations observed in micro-quasars \citep[\eg,][]{RemMcC:2006}.  We note
that, because self-lensing is more probable for the higher mass binaries,
which generate a stronger PTA signal, preferential removal of the high mass
MBHB candidates helps even further to satisfy both the lensing and the PTA
constraints. Future constraints using both the GWB and lensing
statistics will help to vet the population of EM-identified MBHBs.

Even if all of the known MBHB candidates are real, and consistent with GWB
measurements, the \cite{Graham+2015b} and \cite{Charisi+2016} studies may not
have found the MBHBs which exhibit periodic flaring. This is because the
selection processes used to identify periodic light curves preferentially pick
out sinusoids. Lensing events in the \cite{Graham+2015b} and
\cite{Charisi+2016} samples may have simply been discarded due to this bias.
Hence, we point out that it is crucial that present and future time-domain
surveys search for periodicity that may not necessarily be sinusoidal, but
spiky as in the case of the self-lensing flares calculated here.

Finally we note that some lensing flares may be at too low of a 
magnification to be picked out from the data by eye. This could be due to flux
contributions from components of the MBHB system that are not being lensed.
Exploration of this possibility requires a more careful search in the existing
MBHB candidate light curves. We leave this for a separate study.

The probability calculation of Eq. (\ref{Eq:Prob}), plotted in Figure
\ref{Fig:PsandTs}, assumes a flat prior on the inclination of the binary orbit
to the line of sight. If, however, a set of MBHB candidates is selected by
identification of periodicity in AGN light curves due to the relativistic
Doppler boost, the conditional probability of observing a lensing event
increases. This is because observation of the Doppler boost restricts the
range of possible binary inclinations.

To compute this conditional probability of lensing given an observed
Doppler-boost, we compute the range of orbital inclinations available to a
binary given that it has been identified as a Doppler boost candidate. For a
given Doppler-boost MBHB candidate one has a measurement of $n_a$ through the
measured binary orbital period and mass, the amplitude of variability
$A_{\rm{obs}}$, and the spectral slope in the observing band $\alpha$. The
binary mass ratio and orbital inclination are unknown but are related to
eachother and the measured parameters by, 
\begin{equation} 
I(q) = \cos^{-1}{\left[ (1+q) \frac{A_{\rm{obs}}}{3 - \alpha} \sqrt{n_a} \right]},
\end{equation} 
to order $v/c$ \citep[see][]{PG1302Nature:2015b} and where we have assumed
that only the secondary is emitting light. For measured values of $n_a$,
$\alpha$, and $A_{\rm{obs}}$, we can calculate the range of possible binary
orbital inclinations from the range of mass ratios. We require as before that
a significant lensing event be one that brings the source and lens within one
Einstein radius of each other $I \leq \sin^{-1}{\left[ \sqrt{2/n_a}\right]}$
(Eq. \ref{Eq:INE}). But now, instead of comparing this range of inclination
angles to the full unrestricted $\pi/2$ (Eq. \ref{Eq:Prob}), we take into
account a smallest $I(q=1)$ and largest $I(q=0)$ possible binary inclination
set by the Doppler-boost observation. Then the increased probability of
lensing is,
\begin{equation}
\mathcal{P}(L|D) \approx \frac{  \sin^{-1}{\left[ \sqrt{2/n_a}\right]  - \mathcal{R}e\left\{I(q=1) \right\} }  }{ \mathcal{R}e\left\{I(q=0)\right\} - \mathcal{R}e\left\{I(q=1)\right\} }.
\label{Eq:PLD}
\end{equation}

\begin{figure}
\begin{center}$
\begin{array}{c}
\includegraphics[scale=0.31]{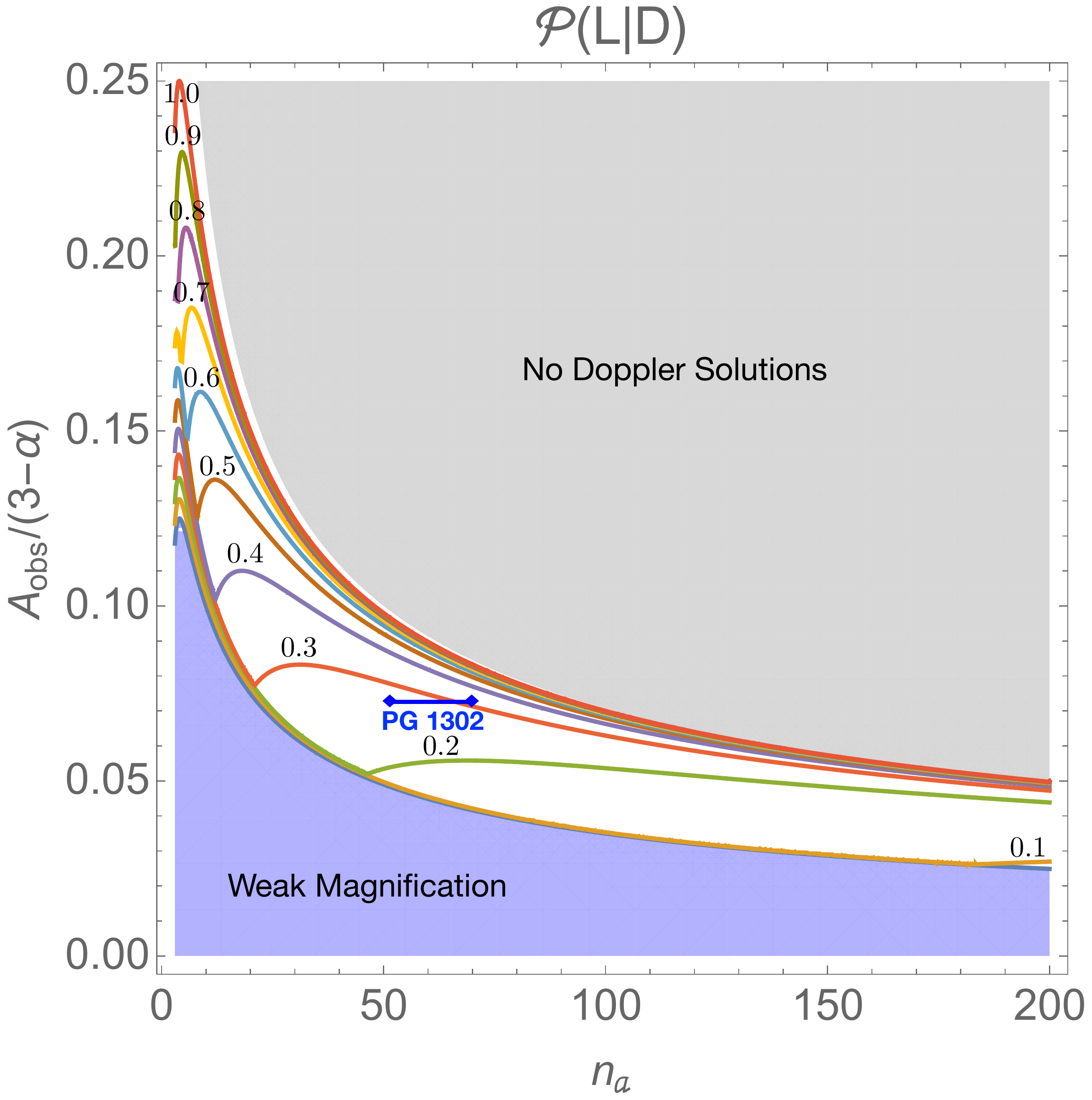} 
\vspace{-0.5cm}
\end{array}$
\end{center}
\caption{
The probability of a lensing event (of the secondary accretion disc)
given that the Doppler boost is observed (labeled contours). The Doppler-boost
signature has modulation amplitude $A_{\rm{obs}} \equiv \Delta
F_{\nu}/F|_{\max}$ for a binary with scaled separation $n_a \equiv a/R_S$ and
observed with spectral index $\alpha$ ($\alpha=2$ here).  The upper-right grey
region labeled `No Doppler Solutions' is where $A_{\rm{obs}}$ is not possible,
even for an edge-on binary. The lower-left blue region labelled `Weak
Magnification' is where the full range of allowed binary inclinations is at a
higher latitude than required for significant lensing magnification.
}
\label{Fig:PDop}
\end{figure}

We plot this probability as a function of binary separation and the
spectral slope dependent Doppler-boost amplitude $A_{\rm{obs}}/(3 - \alpha)$
in Figure \ref{Fig:PDop}. The upper-right gray region of Figure \ref{Fig:PDop}
labeled `No Doppler Solutions' is where the $\alpha$-scaled Doppler-boost
amplitude is not achievable given the binary separation (no matter the orbital
inclination or mass ratio). The lower-left blue region labeled `Weak
Magnification' is where the combination of $A_{\rm{obs}}/(3 - \alpha)$ and
$n_a$ implies a binary inclination that is larger than that required for
significant lensing, for any binary mass ratio. In between these two
disfavoured regions we plot contours of the conditional probability, Eq.
(\ref{Eq:PLD}). As expected, the larger the observed Doppler-boost amplitude,
the more narrow the allowed binary inclinations and the higher the conditional
probability of lensing.

As an example of the utility of this calculation, we plot the
position of the known Doppler-boost MBHB candidate PG 1302 on Figure
\ref{Fig:PDop}. We use the measured values of $\alpha=1.1$ and
$A_{\rm{obs}}=0.14$ in the optical and the range of separations given by the
binary masses consistent with the Doppler-boost scenario $M=10^9-10^{9.4}
\Msun$ \citep{PG1302Nature:2015b}. It is interesting to note that PG 1302
falls in the allowed lensing regime. If the Doppler-boost interpretation is
correct for PG 1302, then there is an $\sim30 \%$ chance for PG 1302 to have
hosted a lensing flare. Because PG 1302 does not exhibit a lensing flare,
however, further analysis could be used to restrict the binary inclination and
mass ratio of the putative MBHB in PG 1302. We leave such an analysis for a
future study.

Because measurements of $n_a$, $A_{\rm{obs}}$, and $\alpha$ exist for the MBHB
candidates plotted in Figure \ref{Fig:PsandTs}, an analysis of the lensing
probability under the assumption of a Doppler boost origin could be readily
examined for the known candidates. We leave this analysis, which should be
accompanied by a more rigorous vetting of the Doppler assumption, for future
work.

Note that this calculation, as well as the reverse calculation: the
probability of observing the Doppler boost given a lensing flare, will depend
on the relative brightness of the emitting regions around each BH and on the
binary mass ratio. That is, the Doppler boost signal should be present when a
lensing signature is observed, but it can be diminished for near a equal-mass,
circular binary where both components are moving at nearly opposite orbital
speeds and are emitting at the same brightness.

\subsection{Probe of binary orbit and accretion}

In the point-source case, observation of one or both lensing flares
could exquisitely constrain the binary orbital inclination and mass ratio, two
parameters which are otherwise difficult to constrain
\citep[e.g.][]{PG1302Nature:2015b,Kun+2015}. As seen in Figure
\ref{Fig:PS_LCs}, the magnification of either secondary or primary flare is
very sensitive to the orbital inclination of the binary. Additionally, a
measurement of the relative magnification of both flares would directly give
the binary mass ratio (see \S \ref{S:FiniteSource}). A non-detection of the
second flare would put an upper limit on the binary mass ratio if the relative
flux of each accretion disc structure can be determined from the primary
flare.

As we have shown in \S\ref{S:FiniteSource}, when the finite size of the lensed
accretion disc becomes comparable to the Einstein radius of the lens, the
lensing flare will have temporal structure that depends on the wavelength.
Optical/IR radiation probes the cooler, outer regions of the disc, while
shorter wavelength, UV to X-ray radiation probes the very inner regions of the
disc where the point-source approximation is valid. Similar lensing tomography
has been carried out to probe the accretion disc structure of multiply imaged
quasars that suffer micro-lensing by stars in the lens galaxy
\citep[\textit{e.g.}][]{Kochanek+2007, Dai+2010, Jimenez+2015, Chartas+2016}.
In the case of multiply images quasars, the source structure is probed by
comparing changes in the flux ratios of multiple images due to an unknown
distribution of stellar lenses. In contrast, the binary self-lensing technique
has the advantage that the distance from source to lens, as well as the mass
and trajectory of the lens, could be known with higher certainty.  Also, 
self-lensing probes specifically the accretion discs around BHs in binaries, rather
than the accretion discs around single BHs.

Observations of the lensing of both primary and secondary discs would allow us
to separately study the contribution from both mini-discs. By elimination, the
contribution from an outer circumbinary disc could also be identified. This
would be invaluable to testing and informing theoretical work on the spectral
signatures of circumbinary accretion discs \citep[\eg,][]{Farris:2015:Cool,
GenerozovHaiman:2014, Roedig+Krolik+Miller2014}, and circumbinary accretion in
general.
        
If the source periodicity is overwhelmed by accretion disc variability rather
than the Doppler boost, then the flare will not necessarily occur at the
relative phase displayed in Figures \ref{Fig:PS_LCs}, at the points of average
brightness. However, a repeating flare on top of a periodic light curve that
does not occur at the expected Doppler boost phase will teach us when in the
orbital phase accretion occurs on to the lensed BH, especially if a weaker
Doppler-boost signal can be recovered for phase reference. This could allow us
to test hydrodynamical models of accretion on to MBHBs, especially if multiple
periods in the light curve are detected \citep[\eg,][]{DHM:2013:MNRAS,
Farris:2014, PG1302MNRAS:2015a, PG1302-Maria}.

If the Doppler variability is detected, the relative phase of the flare can be
fixed and can act as an excellent tracer of binary orbital dynamics. If  the
discs surrounding the MBHs are very massive, they could cause Newtonian
precession of the orbit, or if the binary is compact enough (and
eccentric), relativistic precession effects will become important. Orbital
precession will cause the timing (pericentre precession) and magnitude
(apsidal precession) of the flare to change, but the flare will always appear
at the same relative phase in the light curve. Furthermore, the shape of the
Doppler light curve will encode the orbital elements of the binary (\eg, only
circular orbits considered here are sinusoidal). This scenario could
allow high precision measurements of precession effects.

Because the lensing flare reliably tracks the orbit of the binary and will be
common for binaries near merger (see Figure \ref{Fig:PsandTs}), the existence of
a lensing flare would also facilitate attempts to track an EM chirp
accompanying the GW chirp of a MBHB at merger\footnote{Or a stellar mass BH+BH
binary, or a BH+neutron star binary at merger if one of the binary components
can emit bright EM emission \citep{Schnittman+2017}.} \citep{Haiman:2017}.

\subsection{Extensions}
\label{S:Extensions}

Our calculations should apply to a wide range of real MBHBs. They do not,
however, cover all possibilities. Below we list a set of possible extensions
of this work that may apply to some MBHBs. These extensions would allow
predictions which include lensing effects to be compared with observations of
even more systems.

\begin{enumerate}
    \item 
    We have not included the finite light travel time from the source accretion
    disc to the observer at different phases of the orbit. This will contribute an
    $\mathcal{O}(v_{||}/c)$ effect to the Doppler plus lensing light curves,
    where $v_{||}$is the line-of-sight orbital velocity of the source.

    \item 
    Accretion disc variability (both stochastic and periodic components) could
    cause the flare magnitude to vary from orbit to orbit, but this can be
    tracked from the continuum light curve. Future work should also consider
    the amplitude of stochastic AGN variability relative to expected lensing
    magnification.

    \item When considering finite-sized source effects we have not included the
    strong-field lensing of the source accretion disc by the source BH, which
    will alter the shape of the source for nearly edge-on accretion disc
    inclinations.

    \item 
    We have assumed that the binary is on a circular orbit throughout. 
    If significant eccentricity exists in the orbit, the time-scale for the 
    lensing flare will depend on the argument of pericentre.

    \item 
    We have not fully considered the much more short-lived regime where the
    BHs are within 10s of Schwarzschild radii of each other and relativistic
    ray tracing must be performed rather than the point-mass lens
    approximation we make here. Future work should consider this case: while
    the residence time of the binary is short in this regime, the fraction of
    inclination angles for which strong-field lensing occurs approaches unity
    as the binary orbital separation approaches two Schwarzschild radii. If
    gas can follow close to merger, this lensing would be an almost necessary
    component of the electromagnetic counterparts to the GW sources of 
    space-based gravitational wave detectors (LISA/PTAs)
    \citep[see,][]{Haiman:2017}. Such calculations are important for future
    all-sky surveys \citep[\eg, the Large Synaptic Survey
    Telescope][]{LSST} and \citep[  the Zwicky Transient
    Facility][]{ZWT:2014}, which will monitor a larger number of AGN allowing
    the possibility of detecting even these rarest, compact MBHBs.
     
    \item If the lensing BH has an optically thick accretion flow around it, with
    extent greater than its Einstein radius, the lensing flare would be
    blocked, but in this case we would expect an eclipse in place of the
    lensing flare with time-scale indicative of the primary disc size. If there
    is lensing of the primary accretion disc in the same system, eclipsing by
    and lensing of the primary accretion disc could provide independent
    constraints on the accretion disc size and opacity. 

    \item 
    It is not clear whether or not a circumbinary disc will
    be aligned with the orbital plane of the binary 
    \citep[e.g.][]{Miller+2013, AlyNixon_Misalign+2015}. If the
    binary orbit is co-planer with a circumbinary disc, lensing flares could be
    obscured entirely. However, likely so would any other periodic emission from
    the binary mini-discs or circumbinary disc inner edge. In this case, the system
    would simply not be identified as a MBHB candidate via periodicity. It may be
    possible, however, to detect such periodically changing flux in the 
    time-dependent broad line emission.

    \item 
    If there is a correlation between the binary angular momentum axis and the
    inclination of a surrounding dust torus, then it may be that the largest
    separation binaries, for which the binary inclination must be closest to
    edge on for significant lensing to occur, will be preferentially obscured.
    A lack of lensing events in confirmed MBHB candidates could even
    hint towards this scenario in the future. However, this would mean that
    closer to face-on systems could exhibit lensing flares in the IR, due to
    dust reverberation \citep[\eg,][]{DZ:2017}, such lensing echoes are the
    subject of future work.

\end{enumerate}

\section*{Acknowledgements}
The authors thank Alberto Sesana, Maria Charisi, James Guillochon, Zolt\'an
Haiman, Atish Kamble, Bence Kocsis, Avi Loeb, Frits Paerels and Roman Rafikov
for useful comments and discussions. The authors thank the anonymous referee
for comments that improved the clarity of this manuscript. Financial support
was provided from NASA through Einstein Postdoctoral Fellowship award number
PF6-170151 (DJD) and also through the Smithsonian Institution and the Sprague
Foundation (RD).

\bibliographystyle{mnras}
\bibliography{refs}
\end{document}